\documentclass{aa}
\usepackage{graphics,txfonts}

\newcommand{\hr}{HR\,3831}
\newcommand{\cir}{$\alpha$\,Cir}
\newcommand{\equ}{$\gamma$\,Equ}
\newcommand{\ms}{m\,s$^{-1}$}
\newcommand{\kms}{km\,s$^{-1}$}

\newcommand{\tefv}[1]{$T_{\rm eff}={#1}$~K}

\newcommand{\lggv}[1]{$\log\,{g}={#1}$}

\newcommand{\ha}{H$\alpha$}

\newcommand{\nd}{\ion{Nd}{iii}}
\newcommand{\pr}{\ion{Pr}{iii}}
\newcommand{\ndlin}{\ion{Nd}{iii} 6145.07~\AA}
\newcommand{\nur}{\nu_{\rm rot}}
\newcommand{\wl}{$W_\lambda$}
\newcommand{\va}{$\langle V \rangle$}
\newcommand{\vb}{$\langle V^2 \rangle$}
\newcommand{\vc}{$\langle V^3 \rangle$}

\newcommand{\fifps}[2]{\centering\resizebox{#1}{!}{\includegraphics{#2}}}

\newcommand{\figps}[1]{\resizebox{\hsize}{!}{\rotatebox{0}{\includegraphics{#1}}}}

\newcommand{\beq}{\begin{equation}}
\newcommand{\eeq}{\end{equation}}

\begin{document}

\title{Pulsational line profile variation of the roAp star \hr
\thanks{Based on observations obtained at the European Southern Observatory, La Silla, 
Chile (ESO program No. 66.D-0241A).}}

\author{O. Kochukhov}

\institute{Department of Astronomy and Space Physics, Uppsala University, SE-751 20, Uppsala, Sweden\\
           \email{oleg@astro.uu.se}}

\date{Received 30 April 2005 / Accepted 12 September 2005}

\abstract{
We report the first comprehensive investigation of the line profile variation caused by
non-radial pulsation in a magnetic oscillating chemically peculiar star. Spectrum
variation of the well-known roAp star \hr\ is detected using very high-resolution high
signal-to-noise spectroscopic time-series observations and are followed through the whole
rotation cycle of the star. In contrast to previous photometric pulsational measurements of
roAp stars, our spectroscopic observational material admits straightforward astrophysical
interpretation and, hence, opens new exciting possibilities for direct and accurate analysis of
the roAp pulsations. We confirm outstanding diversity of pulsational
behaviour of different lines in the \hr\ spectrum and attribute this phenomenon to an interplay
between extreme vertical chemical inhomogeneity of the \hr\ atmosphere and a running pulsation
wave, propagating towards the upper photospheric layers with increasing amplitude. 
The rapid profile variation of the \nd\ 6145~\AA\ line, which shows the maximum pulsational
disturbances in the studied wavelength region, is characterized by measuring changes of its
equivalent width and the first three moments. Each of these observables exhibit almost purely
sinusoidal variation through the pulsation cycle, with the amplitude and phase clearly
modulated by the stellar rotation. We demonstrate that rotational modulation of the radial
velocity oscillations cannot be fully explained by an oblique axisymmetric dipole ($\ell=1$,
$m=0$) mode, implied by the classical oblique pulsator model of roAp stars. 
Pulsational variation of the higher moments, in particular the line width, reveal substantial
contribution of the high-order ($\ell=3$) spherical harmonics which appear due to distortion of
pulsations in \hr\ by the global magnetic field. We interpret observations with a novel numerical 
model of the pulsational variation and rotational modulation of the line profile moments in roAp 
stars. The comparison between observed and computed amplitudes and phases of the radial velocity 
and line width variation is used to establish parameters of the oblique
pulsator model of \hr. Furthermore, definite detection of pulsational variation in lines of light
and iron-peak elements enables the first 3-D mapping of pulsations in non-radially oscillating
star.
}

\maketitle

\section{Introduction}
\label{intro}

\subsection{RoAp stars}

The rapidly oscillating Ap (roAp) stars are cool magnetic chemically peculiar stars,
pulsating in high-overtone non-radial acoustic {\it p-}modes. There are 34 roAp stars known at
present time (Kurtz \& Martinez \cite{KM00} for a recent review). These objects oscillate with periods in the
range of 6--21~min (Kurtz \& Martinez \cite{KM00}; Elkin et al. \cite{ERC05}), 
while their light variation amplitudes rarely exceed 10~mmag in
Johnson $B$. Pulsations in roAp stars bear certain resemblance to the solar acoustic {\it
p-}mode oscillations, but are characterized by smaller number of excited modes and have 
three order of  magnitude larger amplitudes compared to the {\it p-}mode oscillations in
the Sun. Photometric investigations of roAp stars were carried out during the last 25
years and have yielded unique asteroseismic information on the internal structure and
fundamental parameters of roAp pulsators (e.g., Matthews et al. \cite{MKM99}; Cunha et al.
\cite{CFM03}).

Since the discovery of roAp pulsations by Kurtz (\cite{K78}) it became clear that 
strong magnetic fields in these stars have a defining role in exciting the oscillations and
shaping the main pulsation properties. It was found that the amplitude and phase of the rapid
light variation are modulated with the stellar rotation and that the phases of the
magnetic field and pulsation amplitude extrema typically coincide with each other (Kurtz \cite{K82}).
Detailed frequency analyses of the roAp photometric light curves convincingly demonstrated
that main characteristics of their rapid brightness variation can be attributed to the oblique
axisymmetric dipole ($\ell=1$, $m=0$) modes, aligned with the axis of nearly axisymmetric
quasi-dipolar magnetic field. This {\it oblique pulsator model}, first proposed by Kurtz
(\cite{K82}), gave rather successful and straightforward phenomenological explanation of the
main features in the roAp frequency spectra. However, subsequent studies of roAp
pulsations revealed that the mode geometry in some stars defied a simple interpretation in
terms of a single spherical harmonic (e.g., Kurtz et al. \cite{KMM89}). Moreover, gradual
accumulation of the photometric information about pulsations in some best-studied monoperiodic
roAp stars with oblique dipole modes provided evidence for significant deviations of the mode
geometry from the expected axisymmetric dipole shape -- an effect loosely called ``distorted
dipole modes'' (Kurtz et al. \cite{KWR97}).

Several theoretical investigations (e.g., Dziembowski \& Goode \cite{DG85}; Shibahashi \&
Takata \cite{ST93}; Cunha \& Gough \cite{CG00}) have argued that combined effects of the magnetic
field and rotation may be responsible for the departure of the surface geometry of
pulsational perturbations from a pure spherical harmonic. In fact, one of the most recent 
theoretical studies, presented by Bigot \& Dziembowski (\cite{BD02}), has challenged the
general framework of the classical oblique pulsator model of Kurtz (\cite{K82}) by showing
that accurate non-perturbative treatment of the interaction between stellar rotation,
pulsations, and magnetic field may lead to an entirely new geometrical picture of the roAp
oscillations. Calculations by Bigot \& Dziembowski (\cite{BD02}) suggested that pulsation
eigenmodes are represented by a complex superposition of spherical harmonic functions and
may contain substantial  non-axisymmetric components. Furthermore, strong effects due to
the centrifugal force are expected to break the locking of the pulsation axis with the dipolar
magnetic field, so that pulsations are aligned with neither the magnetic nor rotation axes.
It should be noted, however, that calculations of Bigot \& Dziembowski (\cite{BD02}) are
limited to the {\it p-}mode pulsations in stars with magnetic field strength $\la$1~kG and
are not directly applicable to the majority of roAp pulsators, which have surface magnetic fields
of a few kG.

In contrast to the approach followed by Bigot \& Dziembowski (\cite{BD02}), Saio \& Gautschy
(\cite{SG04}), and Saio (\cite{S05}) studied the effects of strong magnetic field on axisymmetric
{\it p-}mode oscillations assuming an alignment between pulsations and dipolar magnetic field and
disregarding distortion of pulsation modes by the stellar rotation. Saio \& Gautschy (\cite{SG04})
find substantial deviation of the surface structure of pulsational fluctuations from a single
spherical harmonic. In particular, the $\ell=1$ mode is distorted by a dipolar magnetic field in
such a way that pulsations are described by a superposition of the spherical harmonics with odd
angular degree $\ell$ and pulsation amplitude is strongly confined to the magnetic axis in the
outer stellar layers.

In the view of the diversity and increasing sophistication of the theoretical speculations about
the geometry and physics of roAp oscillations it becomes apparent that modelling of the photometric
light curves of roAp stars is generally insufficient for accurate determination of the structure of
pulsation modes and is unable to provide useful tests of theoretical predictions. Information
content of the high-speed photometric observations is small due to averaging of pulsational
disturbances over the visible stellar hemisphere (see Saio \& Gautschy \cite{SG04}) and highly
uncertain because rapid light variation in roAp stars involves non-linear and non-adiabatic effects
which are not at all  understood (Watson \cite{W88}; Medupe \cite{M02}; Saio \cite{S05}). It has
been argued (e.g., Kurtz \cite{K90}) that spectroscopic time-resolved observations, in particular
investigations of pulsational variation in the Doppler-broadened spectral line profiles at
different rotation phases, can potentially provide much more precise and unprecedentedly complete
information about the vertical and horizontal structure of {\it p-}modes and its relation to the
magnetic field topology, chemical inhomogeneities, and anomalous atmospheric structure of Ap stars.
Such spectroscopic studies will eventually allow us to address the underlying questions about the
physics of pulsating peculiar stars, for instance, the problem of excitation of the magnetoacoustic
oscillations or nature of the mode selection mechanism.

\subsection{Spectroscopic studies of the roAp pulsations}

Until recently our knowledge of the {\it p-}modes oscillations in roAp stars was based essentially
on the high-speed photometric observations. Early time-resolved spectroscopic studies (Matthews et
al. \cite{MWW88}; Libbrecht \cite{L88}) were focused on the detection of pulsational radial
velocity (RV) changes and deriving the ratio of RV and photometric amplitude. Subsequent more
detailed low spectral resolution investigations of \cir\ and \hr\ by Baldry et al.
(\cite{BKB98,BBV98}) revealed unexpected diversity of the pulsational amplitude and phase in
different wavelength regions, showing that for roAp stars the concept of a ``typical''
amplitude of RV oscillations is meaningless. Baldry et al. (\cite{BVB99}) and Baldry \& Bedding
(\cite{BB00}) also found systematic change of pulsational amplitude and phase in the wings of the
hydrogen \ha\ line, which was suggested to originate from a strong depth-dependence of the
pulsational characteristics in roAp atmospheres.

High-resolution observations of \equ\ by Kanaan \& Hatzes (\cite{KH98}) localized RV
variations of individual metal lines and confirmed that pulsations change dramatically from
one spectral line to another. This phenomenon remained a mystery however, because no clear
underlying systematic trends or dependencies were identified. A breakthrough discovery was
made by Savanov et al. (\cite{SMR99}), who demonstrated that in \equ\ the largest
pulsational RV shifts of up to 1~\kms\ occur in the lines of \pr\ and \nd. Follow up very
high spectral resolution time-resolved observations of this star by Kochukhov \&
Ryabchikova (\cite{KR01a}) confirmed these results and also revealed the phase
shifts between variation of singly and doubly ionized lines of the rare-earth elements
(REE). Our study emphasized a stunning discrepancy (sometimes up to two orders of magnitude
in RV amplitude!) between strong pulsations in REE lines and the absence of changes in the
lines of light and iron-peak elements, which do not show oscillations above few tens of
\ms. Discovery of the pulsational spectroscopic variability in 10\,Aql (Kochukhov et al.
\cite{KLR02}) and first analyses of the variation of individual metal lines in \cir\ and 
\hr\ (Kochukhov \& Ryabchikova \cite{KR01b}) showed that this unusual
pulsational behaviour, dominated by the REE lines, is not limited to \equ, but is present
in a very similar form in other roAp stars. This finding was further strengthened by the wide
wavelength region analyses of HR\,1217 (Balona \& Zima \cite{BZ02}), \hr\ (Balona
\cite{B02}), \cir\ (Balona \& Laney \cite{BL03}), 33\,Lib (Mkrtichian et al. \cite{MHK03}), 
and HD\,166473 (Kurtz et al. \cite{KEM03}). 

A plausible explanation of the nature of pulsational spectroscopic variation of roAp stars was
suggested by Ryabchikova et al. (\cite{RPK02}). This study of the atmospheric structure of
\equ\ found compelling evidence for the strong vertical separation (stratification) of chemical
elements. Under the influence of radiative diffusion light and iron-peak elements are
accumulated at the bottom of the \equ\ atmosphere, whereas REEs are pushed high above normal
line-forming region. Ryabchikova et al. (\cite{RPK02}) proposed that the atmospheric layers
contributing to the absorption in REE lines are located substantially higher compared to the
formation depths of other elements. The most recent detailed non-LTE analysis by Mashonkina
et al. (\cite{MRR05}) and Ryabchikova (\cite{R05}) indicates that the \ion{Nd}{ii} and \nd\
lines in \equ\ originate from the optical depths 10$^{-4}$--10$^{-5}$. These high atmospheric
layers rich in REE are characterized by enhanced amplitude of non-radial
pulsations. Within the framework of this model, the phase shifts between variation of different
lines are attributed to a running magneto-acoustic wave propagating outwards in the stellar
atmosphere and increasing in amplitude with height. Similarities in spectroscopic pulsational
behaviour of different roAp stars suggest that they share similar vertical chemical and
pulsational profiles. The method of {\em tomographic mapping} of the vertical structure of roAp
pulsation modes, pioneered by Ryabchikova et al. (\cite{RPK02}) and Kochukhov (\cite{K03}),
uses vertical chemical inhomogeneities as spatial depth filters and aims at reconstructing detailed
vertical cross-section of pulsational perturbations in roAp atmospheres, as can be done for no
other type of main sequence pulsating star.

Despite a dramatic recent progress in our understanding of the vertical structure of the roAp
pulsation modes and the interplay between pulsations and vertical stratification of chemical
elements, relatively little attention has been paid to the problem of inferring the horizontal
geometry of pulsations in roAp stars.  The question of systematic mode
identification, central to the studies of other types of pulsating stars, has not been
thoroughly investigated in the case of magnetic pulsators. To a large extent such an
unsatisfactory situation is explained by the absence of suitable observational material.
Studies of the vertical structure of pulsating cavity capitalize on the extreme vertical
inhomogeneity of the roAp atmospheres and typically make use of a relatively low $S/N$
spectroscopic time series, sufficient just to achieve a detection of pulsations and merely deduce
amplitudes and phases of the RV changes in individual spectral lines. In contrast, probing
the horizontal structure of pulsation modes is necessarily based on the observations of the
pulsational {\it line profile variation} (LPV) and, therefore, requires spectroscopic data of
outstanding quality. Such observations have been presented so far only for one roAp star, \equ\
(Kochukhov \& Ryabchikova \cite{KR01a}). In the study of this object we investigated
variability of the \pr\ 6160~\AA\ and \nd\ 6145~\AA\  lines and probed the horizontal 
structure of the magnetoacoustic {\it p-}mode oscillation using several mode
identification methods. Most
importantly, it was convincingly shown that the moment variation of the REE lines is inconsistent
with a picture expected for axisymmetric ($m=0$) modes of low degree. This discovery of {\it
non-axisymmetric pulsations} in a roAp star gave certain credibility to the generalized oblique
pulsator model of Bigot \& Dziembowski (\cite{BD02}) and demonstrated enormous diagnostic potential
of the high-precision time-resolved spectroscopy of roAp stars.

Despite its interesting pulsational properties, \equ\ is not an ideal target for mapping
the horizontal mode structure because its extremely slow rotation does not allow to put useful
constraints on the surface chemical inhomogeneities and explore oscillations from different
geometrical aspects. Additional complication is introduced by multiperiodicity and 
short growth and decay time of the {\it p-}modes in \equ\ (Martinez et al. \cite{MWN96}).
Evidently, a more fruitful investigation of the LPV due to non-radial oscillation can be carried
out for a monoperiodic roAp star with a high amplitude of pulsational spectral variation and a
reasonably short rotation period. The present paper concentrates on such an analysis of the
well-known roAp star \hr\ and presents the first comprehensive study of the pulsational LPV 
in a roAp star.

\subsection{\hr}

\hr\ (HD\,83368, IM Vel) is one of the best studied rapidly oscillating Ap pulsators. This star
is a brighter component in a visual binary with a separation of 3.29\arcsec\ and a  $\Delta
V=2.84$~mag difference in brightness. \hr\ has a visual magnitude of $V=6.25$ (Hurly \& Warner
\cite{HW83}) and belongs to the SrCrEu group of Ap stars (Houck \cite{H78}). The mean
longitudinal magnetic field of \hr\ was measured by Thompson (\cite{T83}) and Mathys
(\cite{M91}). The longitudinal field varies sinusoidally with an amplitude of 737~G and a mean
value close to zero. The rotation period of $P_{\rm rot}=2.851962$~d was derived from the
magnetic measurements.

Spectroscopic studies of \hr\ were conducted by Polosukhina et al. (\cite{PSH00})
and Shavrina et al. (\cite{SPZ01}). They analysed abundances of several chemical elements
and studied in detail strong rotational modulation of the profile of the \ion{Li}{i}
resonance doublet.

Kochukhov et al. (\cite{KDD03,KDP04}) carried out the first abundance Doppler imaging of \hr\
using average spectra from the present study and additional high-resolution observations. We
combined Hipparcos parallax, Str\"omgren and H$\beta$ photometric indices, and results of the
model atmosphere analysis of the hydrogen lines in \hr\ to derive fundamental stellar
parameters:  \tefv{7650}, \lggv{4.2}, $\log L/L_\odot=1.091$,  $R/R_\odot=2.003$, and 
inclination angle of the stellar rotation axis  $i=68.2\degr$. It was found that many lines in
the \hr\ spectrum show noticeable variation with rotation phase, implying that strong
horizontal abundance gradients are present at the stellar surface. In particular, the Doppler
inversions showed Li and Eu to be concentrated in pairs of opposite small spots, located at
the stellar rotational equator, whereas O exhibits an extended overabundance area along the
big circle perpendicular to the rotational equator. In the context of the present study,
behaviour of the doubly ionized Nd and Pr lines is of special interest because these REE lines
turn out to be ideal pulsation indicators and are extensively analysed in the time-resolved
spectra of roAp stars. Investigation by Kochukhov et al. (\cite{KDP04}) demonstrated that \nd\
and \pr\ lines do not show strong profile anomalies except for the rotation phases close to
magnetic crossover ($\varphi=0.25,0.75$), when pronounced doubling is evident in both the Nd
and Pr absorption features (see also Kochukhov \& Ryabchikova \cite{KR01b}). Our Doppler
mapping of \hr\  attributed this phenomenon to a ring of relative REE underabundance -- a
surface distribution just the opposite to that of oxygen. 

The horizontal abundance structures observed in \hr\ are understood to be a result of
the radiatively driven diffusive separation of
chemical species in the presence of a dipolar magnetic field inclined at a large angle
$\beta$ with respect to the rotation axis of the star. This magnetic topology is in perfect
agreement with the parameters of the dipolar model ($\beta=86.8\degr$, $B_{\rm p}=2.49$~kG)
determined by Kochukhov et al. (\cite{KDP04}), but contradicts the very large 11~kG mean
quadratic field measured in \hr\ by Mathys (\cite{M95}). However, the latter measurements are
known to be severely biased by the neglect of  surface chemical inhomogeneities in the
quadratic field diagnostic procedure (see Sect.~4 in Kochukhov et al. \cite{KDP04}).

Non-radial pulsations in \hr\ were discovered and investigated by Kurtz (\cite{K82}).
Subsequent analysis of the pulsational behaviour using photometric data spanning four
observing seasons was presented by Kurtz et al. (\cite{KWR97}). They revealed strong rotational
modulation of pulsations and determined a principal pulsation frequency
$\nu=1428.0091\,\mu$Hz (11.67~min). The power spectrum around the main frequency is described
by a septuplet structure with a separation of the frequency components by $\nur=4058.265$\,nHz,
corresponding to the stellar rotational period $P_{\rm rot}=2.851976\pm0.00003$~d. Pulsational
light variation of \hr\ is clearly non-sinusoidal with the first-harmonic quintuplet and the
second-harmonic triplet unambigously detectable in the Fourier transform of the time-resolved
photometric data.

A preliminary discovery report of the pulsational variations in individual metal spectral lines
in \hr\ was published by Kochukhov \& Ryabchikova (\cite{KR01b}). In this study we demonstrated
that, similar to the pulsations in other roAp stars, the doubly ionized REE lines in \hr\ are
distinguished by high-amplitude RV variation during pulsation cycle of the star. This was
confirmed by Balona (\cite{B02}), who obtained high-resolution time-series observations in a
wide spectral region at several rotation phases of \hr. Balona found that all metal lines with
large pulsational variability are invariably associated with the \pr\ and \nd\ spectral
features. He also detected a strong decrease of the amplitude of the pulsational RV variation 
from \ha\ to the higher member of the Balmer series, which supported the idea that the 
amplitude of non-radial oscillation is a strong function of height in the atmospheres of 
roAp stars.

The quality of the observational material collected by Balona (\cite{B02}) was not sufficient to
detect and study pulsational changes in the profiles of individual spectral lines. Instead
Balona attempted to use co-added cross-correlation function derived from the \nd\ lines for
tentative mode identification. He suggested that the two frequencies detected in the RV
variability can be attributed to the $\ell=2$, $m=0$ and $\ell=2$, $m=-2$ non-radial modes aligned
with the stellar rotation axis. This mode identification remains controversial since it was not
supported by any quantitative analysis (e.g., line profile calculations) and clearly
contradicts the well-known dipole oblique pulsator model required to explain rapid photometric
variability of \hr.

A new pulsation Doppler imaging method of resolving horizontal structure of stellar non-radial
oscillations was developed by Kochukhov (\cite{K04a}). This technique uses high-resolution
observations of LPV and permits direct mapping of the surface pulsation velocity amplitude without
{\it a priori} parameterization of pulsation with spherical harmonic functions. Preliminary results
of the application of this promising method to \hr\ were reported by Kochukhov (\cite{K04b}).
Doppler mapping of pulsations in \hr\ was based on the spectra analysed here and suggested that
distorted dipole oscillation mode is inclined at a large angle with respect to the stellar rotation
axis and is aligned with the dipole component of the stellar magnetic field.

\section{Observations and spectra reduction}
\label{observ}

\begin{table}[!t]
\caption{Observing log of the time-resolved spectroscopic monitoring of \hr. Columns give
the UT date, heliocentric Julian Date for the beginning and end of observation, the duration of 
continuous monitoring in hours, the range of rotation phases, calculated according to the 
ephemeris of Kurtz et al. (\cite{KWR97}), and the number of spectra obtained on each
night. \label{tbl1}}
\begin{tabular}{ccccc}
\hline
\hline
Date       & HJD$-2451900$         & $\Delta T$ & $\varphi$ & $N$ \\
\hline
05/02/2001 & 45.5533--45.8779 & 7.79 & 0.965--0.079 & 300 \\
06/02/2001 & 46.5568--46.8796 & 7.75 & 0.317--0.430 & 300 \\
07/02/2001 & 47.5470--47.8678 & 7.70 & 0.664--0.776 & 300 \\
13/02/2001 & 53.5328--53.8786 & 8.30 & 0.763--0.884 & 320 \\
14/02/2001 & 54.5286--54.8716 & 8.23 & 0.112--0.232 & 320 \\
15/02/2001 & 55.5209--55.8678 & 8.33 & 0.460--0.581 & 320 \\
\hline
\end{tabular}
\end{table}

Very high-resolution spectroscopic time-series observations of \hr\ presented in this paper
were collected in February 2001 with the Very Long Camera of the Coud\'e Echelle Spectrograph
(CES) fibre-linked to the Cassegrain focus of the ESO 3.6-m telescope. The spectrograph was
configured to reach a resolving power of $\lambda/\Delta\lambda=123\,500$ with the help of the
medium resolution CES image slicer. The detector, the ESO CCD-61 2K$\times$4K chip, allowed us
to record a 40.4~\AA\ spectral interval centred at $\lambda$~6136~\AA. 

During CES observations we continuously monitored \hr\ for about 8 hours on each of the 6
observing nights with a break of 5 nights in the middle of the run. Such an observing strategy
enabled us to obtain 300--320 stellar spectra each night and ensure a good coverage of the
rotation period of \hr, with phase gaps not exceeding 8\% of the rotation cycle (see a log of
observations in Table~\ref{tbl1}). A total of 1860 spectra was collected for \hr. Observations
were obtained with a 70$^{\rm s}$ exposure time, whereas approximately 20$^{\rm s}$ was spent
on reading the CCD detector and file transfer. The dead time was minimized by
exposing only half of the detector in the cross-dispersion direction and additionally binning
by a factor of four in the same direction. This setup allowed us to observe \hr\ with a 78\%
duty cycle and has proved to be an appropriate compromise between a reasonable sampling of the
pulsation period and reaching a sufficient signal-to-noise ratio in individual spectra. 60$^{\rm
s}$ exposures of the ThAr comparison spectrum were obtained every 1.3--1.4 hours during
observations. 

\begin{figure*}[t]
\figps{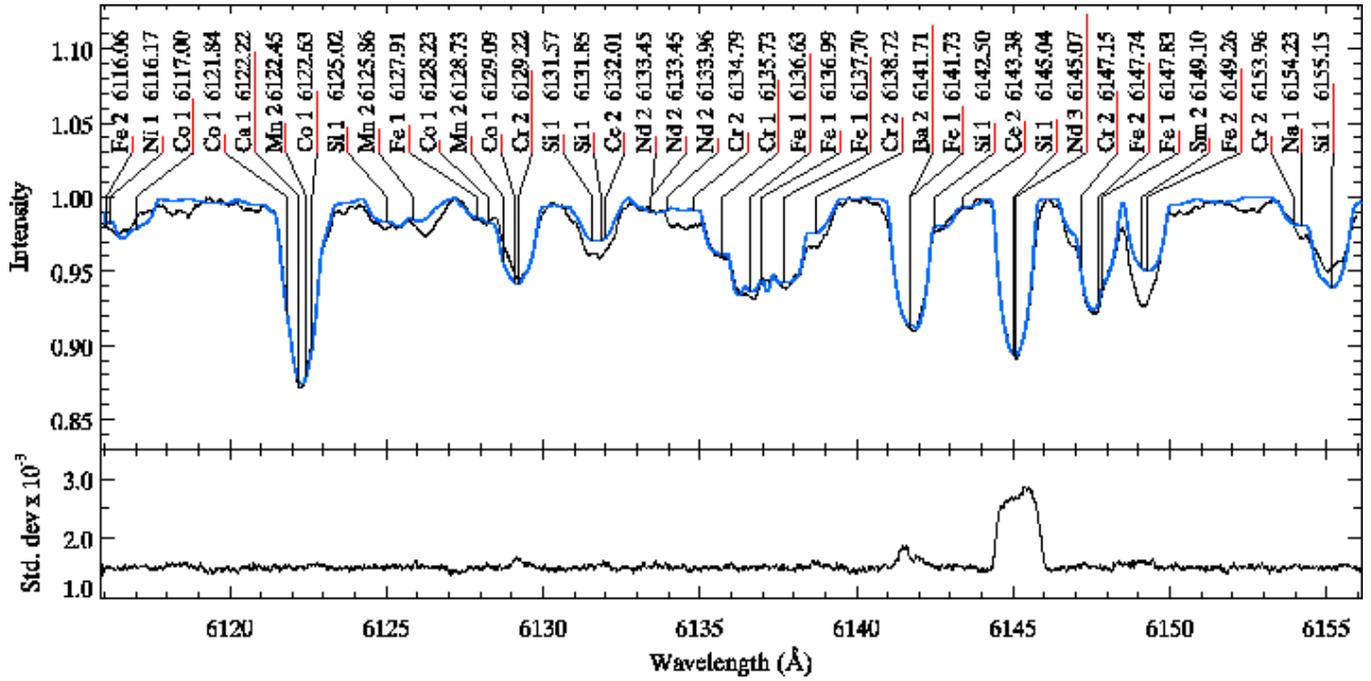}
\caption{
Comparison between the mean spectrum of \hr\ (thin line) and magnetic spectrum synthesis 
calculations (thick line) in the 6116--6156~\AA\ wavelength region is illustrated in the upper panel. 
Identification is given for the strongest spectral lines.
Length of the lines underlining identification is proportional to the strength of the corresponding 
spectral features. The lower panel shows the standard deviation spectrum.}
\label{fig0}
\end{figure*}

A dedicated set of IDL computer routines was developed to perform optimal reduction of the CES
spectra based on the strategy and procedures implemented in the {\sc reduce} package by
Piskunov \& Valenti (\cite{PV02}). The reduction steps included construction of the master flat
field and bias frames, subtraction of bias and correction of the target spectra by the
normalized flat field, subtraction of background, extraction of 1D spectra, and correction for
the overall curvature of the spectra in the dispersion direction. Since the CES instrument was
designed to record observations at resolving power of up to $\lambda/\Delta\lambda=235\,000$,
its instrumental profile  is oversampled in the medium resolution mode ($\approx$\,5 pixels per
resolution element).  We took advantage of this situation and resampled all spectra in the
dispersion direction, descreasing the total number of pixels by a factor of two. This procedure
yielded 1D spectra with $S/N$ in the range of 110--150 per pixel.

A special procedure was developed to establish the wavelength scale and to monitor possible 
instrumental drifts of the spectrograph during observing nights. We identified and measured
positions of 23 emission lines in all available ThAr spectra and then fitted the pixel-wavelength 
relation with a third-order polynomial. Sequencies of comparison spectra
obtained on the same nights were analysed simultaneously to derive common higher order
polynomial coefficients and a time-dependent instrumental shift (the lowest order polynomial
coefficient). The final internal accuracy of the wavelength scale established through this
procedure was $\approx$\,$8\times10^{-5}$ (4~\ms). We found that the zero point of the CES
spectrograph exhibited an overall smooth drift of 60--180~\ms\ due to changes of ambient
temperature and pressure during observations of \hr. At the same time, the average shift
between the consecutive ThAr spectra taken 1.3--1.4 hours apart was at the level of 17--35~\ms. This confirms
an excellent short-term stability of the CES instrument and suggests that possible random
instrumental shifts do not exceed $\sim$\,10~\ms. Consequently, it was justified to linearly
interpolate the zero point between the comparison spectra taken before and after groups of
stellar exposures. Finally, the spectra were shifted to the heliocentric reference frame.

In the last reduction stage the individual spectra of \hr\ were normalized to the continuum.
This procedure was carried out in a series of steps in order to ensure a consistent
normalization of all stellar spectra. First, a third order polynomial was fitted through  the
continuum points of the nightly mean spectra. This average continuum profile was applied to
individual observations and was subsequently corrected for each spectrum by the heavily
smoothed ratio of the average and individual spectra. A similar procedure was followed to
achieve a homogeneity of the continuum normalization of the spectra obtained on different nights.

\section{Spectrum synthesis and line identification}
\label{synthesis}

\begin{figure*}[!t]
\resizebox{12cm}{!}{\includegraphics{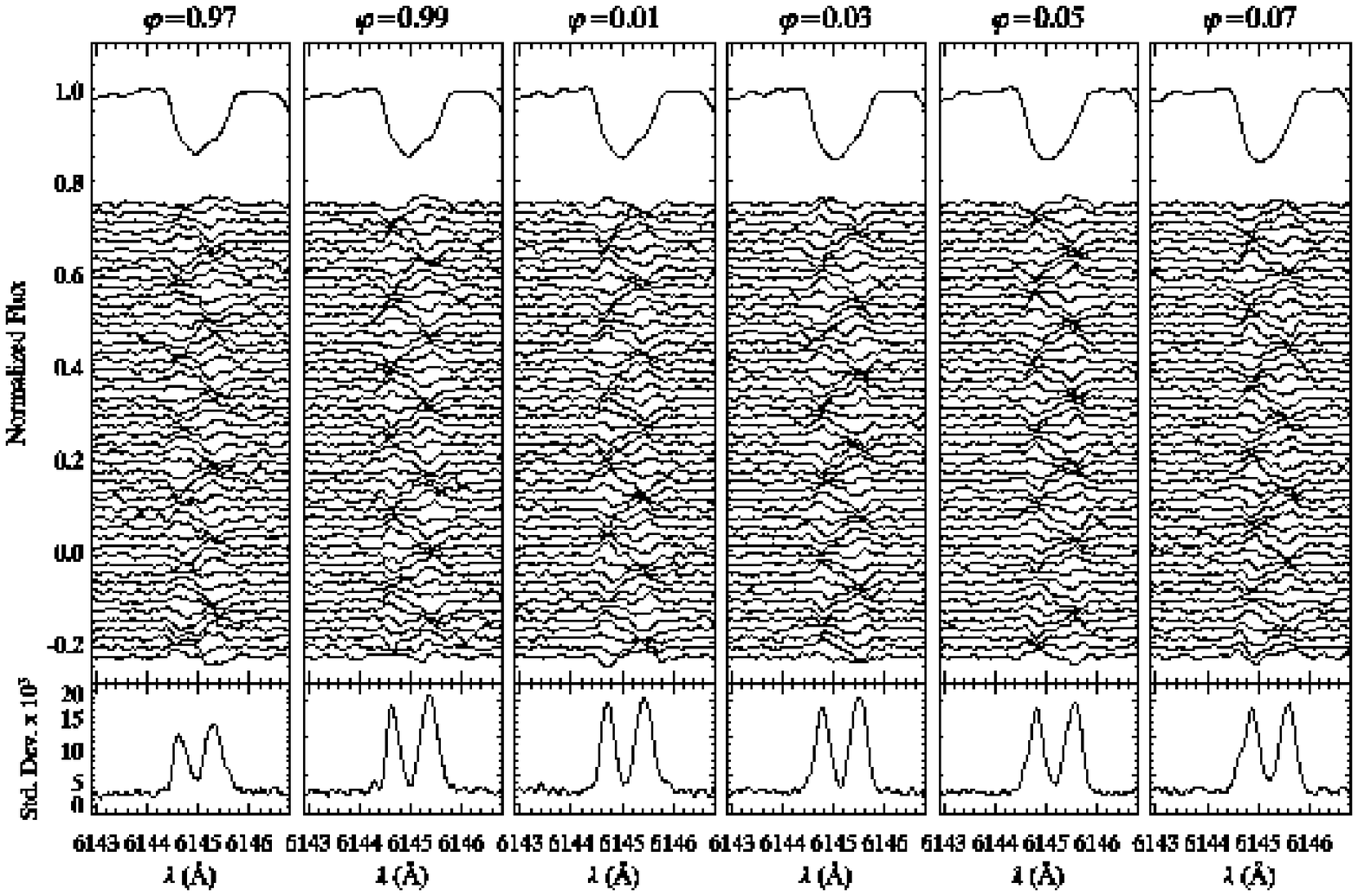}}\\
\resizebox{12cm}{!}{\includegraphics{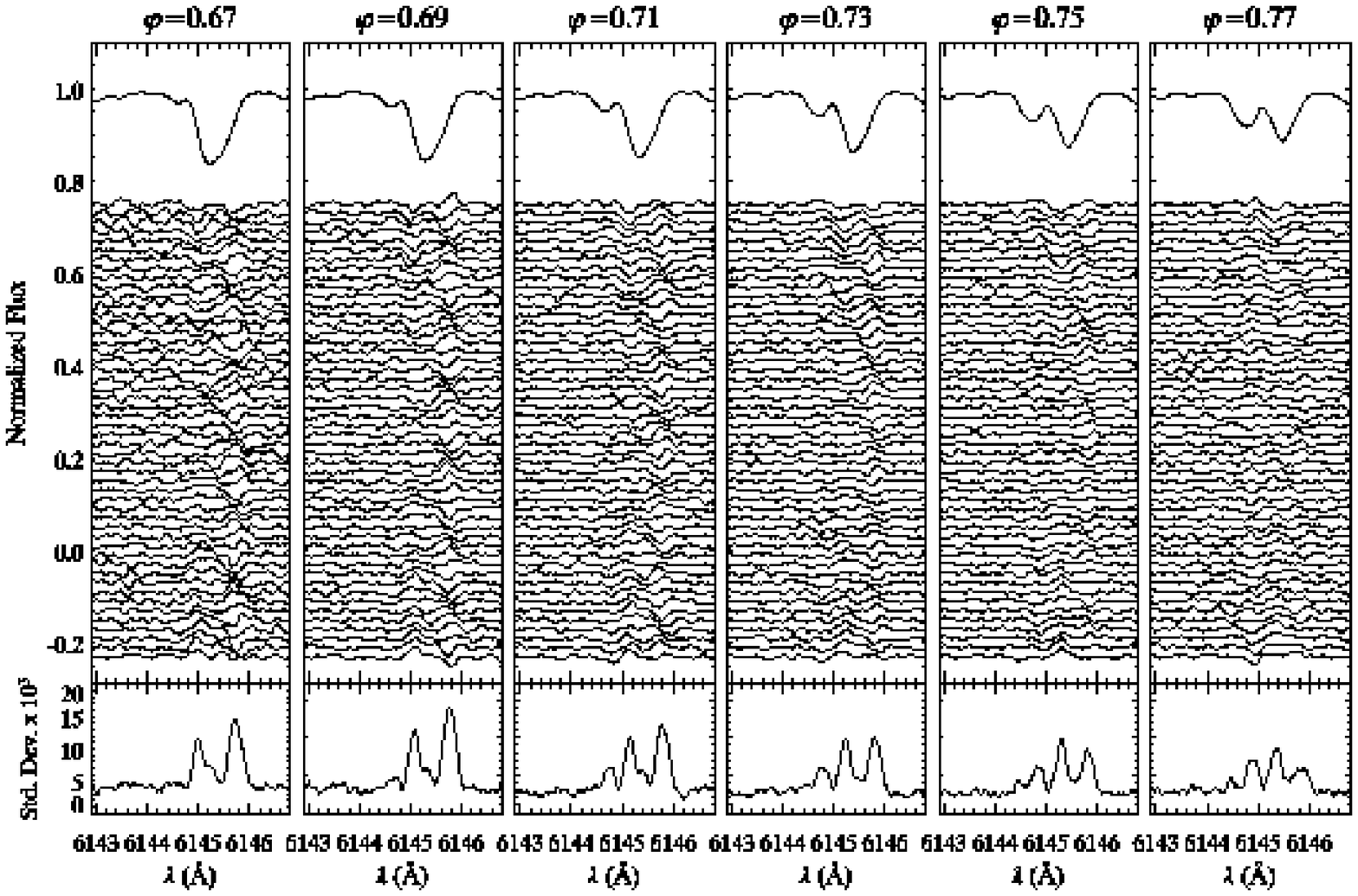}}
\caption{Illustration of rapid variation of the \ndlin\ line in the
spectrum of \hr. The upper panels show LPV recorded on the night of 5
Feb 2001 and the lower panels show observations obtained on 7 Feb 2001. Each of the six horizontal
subpanels presents the average spectrum of \hr\ at the specified rotation phase and 
time sequence of the difference between the average and 50 individual time-resolved observations
obtained during 74~min and covering roughly 6 pulsation periods of \hr.
Consecutive difference spectra are shifted in the vertical direction. The bottom part of 
the horizontal subpanels shows the standard deviation for each pixel of the observed spectra.}
\label{fig1a}
\end{figure*}

\begin{figure*}[!t]
\figps{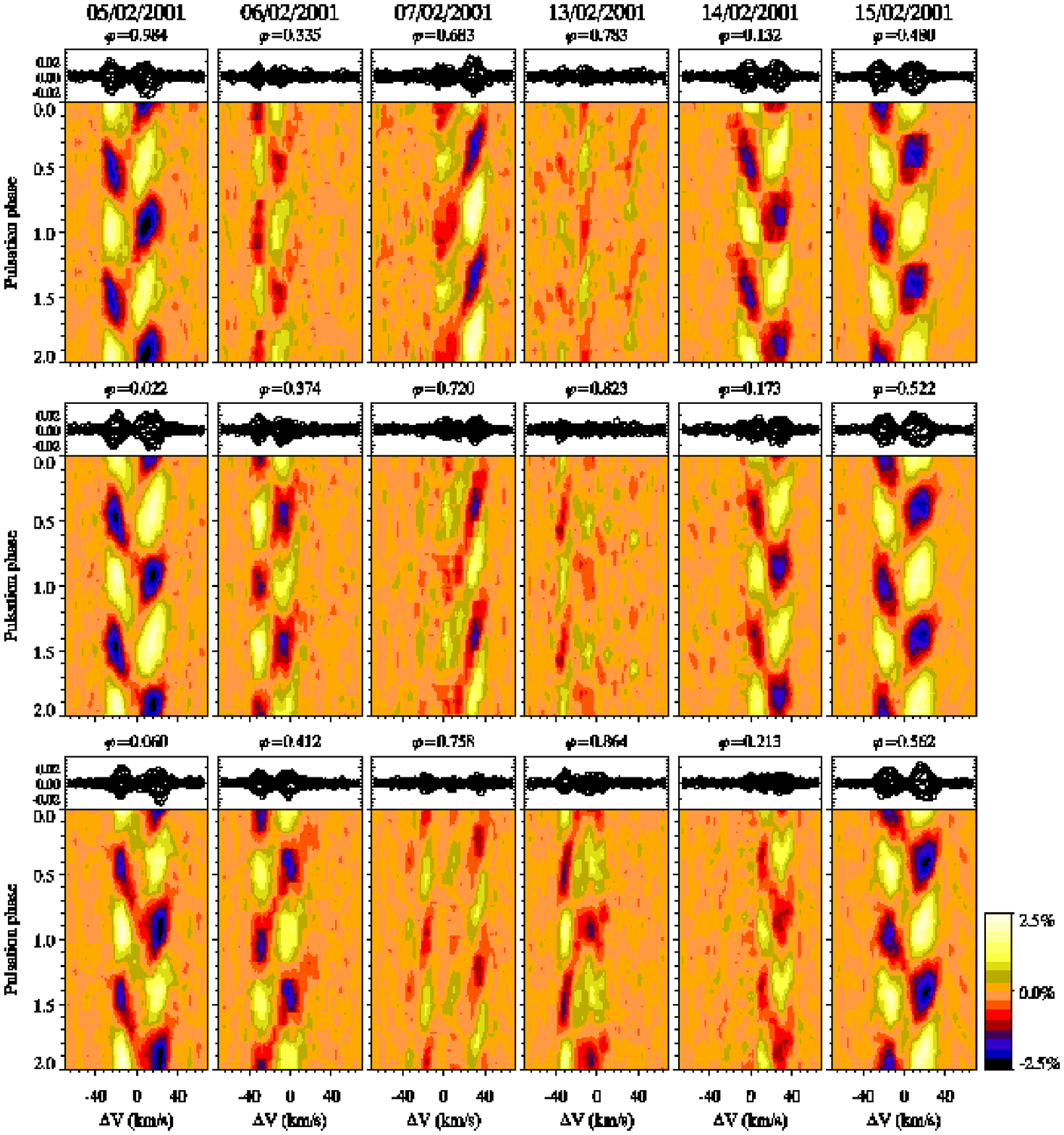}
\caption{
Illustration of the \ndlin\ line profile variation at different rotation
phases of \hr. Each of the greyscale images is based on groups of 23 time-resolved spectra
covering roughly 3 oscillation cycles of \hr\ at specified rotation phases. The upper
plot in each panel shows residuals from the mean line profile. The greyscale plots show
time evolution of the residuals, phased with the main pulsation frequency
$\nu=1427.9958\,\mu$Hz and using the scale of $\pm2.5$\% of the continuum intensity. The
six columns of panels show LPV recorded during six different observing
nights. The UT date is specified at the top of each column. 
Evolution of the variability pattern during observing nights is illustrated by three 
subpanels in each column which display spectra obtained at the beginning, middle, and 
end of the continuous monitoring.
\label{fig1b}}
\end{figure*}

The line identification in the 6116--6156~\AA\ wavelength region of \hr\ was carried out based
on magnetic spectrum synthesis calculations. We adopted the model atmosphere
parameters \tefv{7650}, \lggv{4.2} and the projected rotational velocity $v_{\rm e}\sin
i=33$\,\kms\ determined by Kochukhov et al. (\cite{KDP04}). The relevant parameters of atomic
transitions were extracted from the VALD database (Kupka et al. \cite{VALD}) and were further
fine-tuned by fitting the solar atlas. Synthetic spectra were calculated using the polarized
radiative transfer code \mbox{\sc synthmag} (Piskunov \cite{P99}). A homogeneous magnetic
distribution was adopted and a field strength of $B_{\rm s}=1.75$~kG was estimated from the
dipolar magnetic field model derived for \hr\ by Kochukhov et al. (\cite{KDP04}). 

The comparison between the average spectrum of \hr\ and our theoretical calculation is
illustrated in Fig.\,\ref{fig0}. In general, we find a satisfactory agreement. Remaining
discrepancies are attributed to averaging over spotted surface abundace structures (not accounted
for in our spectrum synthesis) and unidentified spectral lines missing in the theoretical
spectrum. 

In the previous time-resolved observations of the $\lambda$~6150~\AA\ region in slowly rotating
roAp stars (Kochukhov \& Ryabchikova \cite{KR01a}; Kochukhov et al. \cite{KLR02}) we were able
to study several dozen spectral lines. Unfortunately, a moderately rapid rotation of \hr\ makes
a detailed pulsation analysis possible only for a few relatively unblended spectral features
and blends with well-defined components.  Those included the \ion{Ca}{i} 6122.22, \ion{Cr}{ii}
6129.22~\AA\ lines, the blend of \ion{Si}{i} and \ion{Ce}{ii} at $\lambda$~6131.9~\AA, the
\ion{Ba}{ii} 6141.71, \ion{Nd}{iii} 6145.07, \ion{Fe}{ii} 6147.74, and 6149.26~\AA\ lines. Note
that the latter \ion{Fe}{ii} line is severely blended with an unidentified REE line at
$\lambda$~6148.85~\AA\ which shows a high-amplitude pulsation variability in the spectrum of
\equ\ (Kochukhov \& Ryabchikova \cite{KR01a}) and other sharp line roAp stars.

\section{Pulsational behaviour of the \ndlin\ line}

\subsection{Observations of the rapid line profile variability}

The high quality of the spectroscopic observational data obtained for \hr\ enables a direct
detection and comprehensive analysis of the LPV in REE spectral
lines. The strong line of doubly ionized neodymium at $\lambda$~6145.07~\AA\ was found to
exhibit rapid line profile changes which are clearly modulated with the stellar rotation.
Fig.~\ref{fig1a} illustrates variability of this \nd\ line at two representative rotation phases.
At the phase of magnetic extremum, $\varphi=0.02$ (the upper panels in Fig.~\ref{fig1a}), the
average profile of the Nd feature is not significantly distorted by the inhomogeneous surface
distribution of this element. Pulsational variability has a simple symmetric pattern, expected
for the oblique dipole pulsation mode. The line profile as a whole becomes
skewed to the red and then to the blue. This is very different from the pulsational
spectroscopic behaviour of some of the hot non-radial pulsators which is characterized by
high-contrast bumps and dips moving across stellar line profiles (e.g. Schrijvers et al.
\cite{STA04}). Such a behaviour is not observed in \hr\ because of the low angular degree $\ell$
of the excited mode. The dominant pulsational contribution comes from the global $\ell=1$--3
mode which distorts significant part of the line profile, not unlike the spectral changes
typical of the radial stellar pulsation (e.g. Telting et al. \cite{TAM97}).

In contrast to the phase of magnetic extremum, pulsational changes of the \nd\ line at the 
crossover phase (lower panels in Fig.~\ref{fig1a}) have lower amplitude and are more complex.
According to the surface Doppler maps of \hr\ published by Kochukhov et al. (\cite{KDP04}),
rotation phase $\varphi=0.72$ corresponds to a moment when the area of relative Nd
underabundance passes through the centre of the stellar disk. This explains doubling of
the average line profile and suggests that complexity of the oscillation behaviour at this
phase is due to a transition from one pulsation pole to another. Due to a moderately rapid
rotation of the star and the respective substantial Doppler broadening, the areas associated
with both pulsation poles are resolved in the wavelength domain and contribute to
variability in the blue (respectively red) components of the \nd\ line. It is worthwhile to
emphasize that, despite a low amplitude and complexity of the oscillation pattern observed
at the crossover phases, pulsation does not vanish entirely and can still be readily detected
with our high-precision observations.

An alternative representation of the \nd\ line variability is shown in Fig.~\ref{fig1b}. These
greyscale plots encompassing the whole dataset illustrate pulsational changes of the residuals
with respect to the mean spectrum for the beginning, middle, and end of each of the six
observing nights. The maximum deviation from the average spectrum reaches $\pm2.5$\% of the
stellar continuum. Pulsation pattern changes dramatically from one night to another. The data
also show unambiguous evidence for the systematic evolution of the variability pattern during
individual observing nights, capturing rotational change of the aspect angle which becomes
noticeable during a few hours of continuous monitoring (for instance, note an increase of
variability on 6 Feb 2001 and decrease on 14 Feb 2001).

The spectrum variability plots presented in Figs.~\ref{fig1a} and \ref{fig1b} establish a
smooth pattern of the rapid line profile changes in \hr, similar to the pulsational
variability observed in the atmospheres of many other types of stars pulsating with longer
periods and low amplitudes. This observation represents a non-trivial result in the light of
the chemical stratification analysis of Ryabchikova et al. (\cite{RPK02}). They have 
demonstrated that the strong \nd\ lines populating the roAp spectra form in tenuous high
atmospheric layers, whose physical conditions are rather uncertain and where some complex
non-linear pulsation effects may in principle take place. However, our observations do not
provide any evidence for violent phenomena, such as super-sonic shocks or large non-linearity,
associated with the propagation of pulsation waves. Consequently, we believe that, in the
context of the present study, the standard analysis of the pulsational spectroscopic observables,
carried out under (often implicit) assumption that line profile changes are
dominated by linear pulsation velocity fluctuations is fully justified. Nevertheless, it
is useful to keep in mind that the spectroscopic diagnostic explored in our paper is sensitive
to the pulsational perturbations in the upper photosphere and the respective pulsation geometry is not
necessarily the same as for the broad-band luminosity fluctuations originating in the lower
part of the roAp atmospheres.

The data presented here is the first observation of the pulsational LPV in a rapidly rotating roAp star. The
variability pattern found for the \ndlin\ line at the rotation phase of maximum pulsation amplitude
is close to the one expected for an oblique dipolar oscillation (Kochukhov \cite{K04a}). The wave in
the residual spectra propagates symmetrically from the blue to red line wing and then backwards.
This behaviour differs from profile changes of the same \nd\ line in the spectrum of very slowly
rotating roAp star \equ. For the latter object Kochukhov \& Ryabchikova (\cite{KR01a}) discovered asymmetric
variability characterized by the blue-to-red moving features only. However, observations of \hr\ at
the rotation phases of the magnetic crossover respectively minimum pulsation amplitude reveal a
LPV pattern which is more similar to the pulsational changes observed for \equ. Variability is limited to a
narrow wavelength region in one or both wings and the wave in the residual spectra propagates
asymmetrically from blue to red (Fig.\,\ref{fig1b}, 14 Feb) or from red to blue (7 Feb).

\subsection{RV measurements and frequency analysis}

\begin{figure*}[!t]
\fifps{15cm}{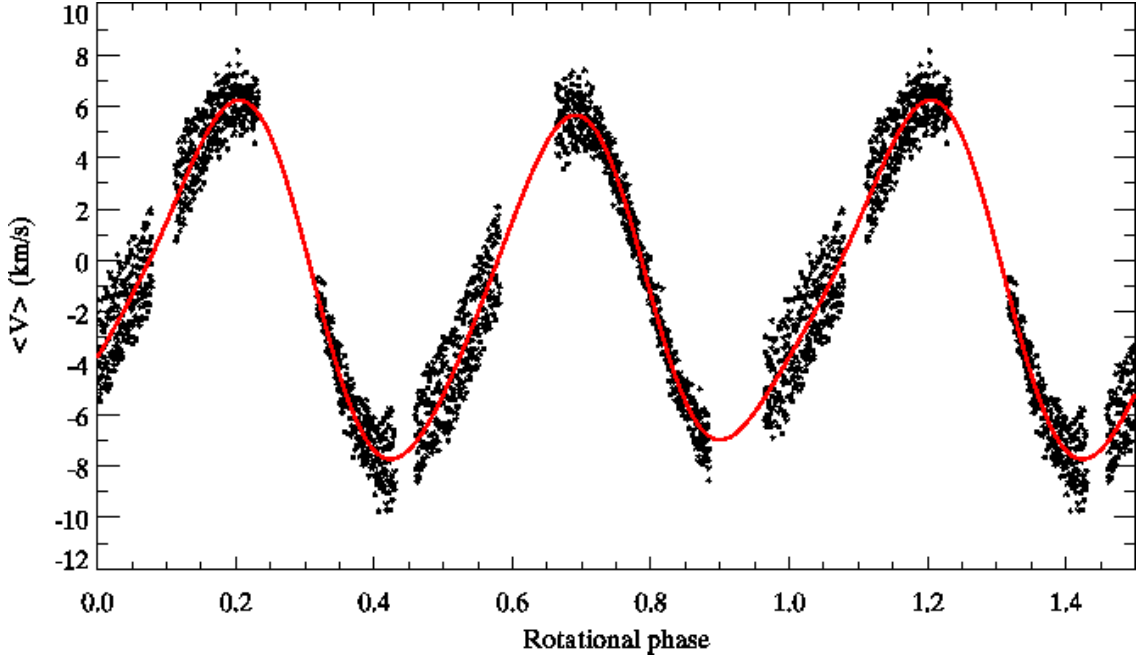}
\caption{Radial velocity variation of the \ndlin\ line in the spectrum of \hr. Symbols show
individual measurements phased with the rotation period $P=2.851976$~d. The smooth curve
illustrates a Fourier fit of the rotational modulation of the average RV.}
\label{fig2}
\end{figure*}

\begin{table*}[!t]
\caption{Results of non-linear least-squares fit of the \ndlin\ line profile moment variations.
\label{tbl2}}
\begin{tabular}{lrcccrcrc}
\hline
\hline
Frequency  & \multicolumn{2}{c}{\wl\ ($\mu$\AA)} & \multicolumn{2}{c}{\va\ (\kms)} & 
             \multicolumn{2}{c}{\vb\ (km$^2$\,s$^{-2}$)} & \multicolumn{2}{c}{\vc\ (km$^3$\,s$^{-3}$)} \\
           & \multicolumn{1}{c}{$A$}  & $\varphi$ & $A$ & $\varphi$ & 
	     \multicolumn{1}{c}{$A$} & $\varphi$ & \multicolumn{1}{c}{$A$} & $\varphi$ \\
\hline
\multicolumn{9}{c}{The fundamental septuplet} \\
$\phantom{2}\nu-3\nur$  & $ 227\pm80$ & $0.242\pm0.056$ & $0.125\pm0.013$ & $0.180\pm0.018$ & $11.07\pm0.39$ & $0.392\pm0.006$ & $120\pm18$ & $0.557\pm0.024$ \\
$\phantom{2}\nu-2\nur$  & $ 175\pm75$ & $0.625\pm0.069$ & $0.012\pm0.016$ & $0.806\pm0.166$ & $ 0.75\pm0.37$ & $0.665\pm0.079$ & $ 56\pm17$ & $0.554\pm0.049$ \\
$\phantom{2}\nu- \nur$  & $2643\pm74$ & $0.926\pm0.004$ & $1.079\pm0.013$ & $0.054\pm0.005$ & $ 8.30\pm0.36$ & $0.312\pm0.007$ & $877\pm17$ & $0.033\pm0.003$ \\
$\phantom{2}\nu      $  & $ 249\pm79$ & $0.420\pm0.051$ & $0.025\pm0.015$ & $0.292\pm0.084$ & $ 0.73\pm0.39$ & $0.029\pm0.084$ & $ 56\pm18$ & $0.174\pm0.052$ \\
$\phantom{2}\nu+ \nur$  & $1887\pm74$ & $0.851\pm0.006$ & $0.594\pm0.014$ & $0.983\pm0.006$ & $22.21\pm0.36$ & $0.723\pm0.003$ & $175\pm17$ & $0.967\pm0.016$ \\
$\phantom{2}\nu+2\nur$  & $ 265\pm75$ & $0.232\pm0.045$ & $0.018\pm0.013$ & $0.550\pm0.165$ & $ 0.93\pm0.37$ & $0.293\pm0.063$ & $ 45\pm17$ & $0.805\pm0.061$ \\
$\phantom{2}\nu+3\nur$  & $  34\pm80$ & $0.132\pm0.378$ & $0.051\pm0.013$ & $0.187\pm0.040$ & $ 3.81\pm0.39$ & $0.804\pm0.016$ & $141\pm18$ & $0.420\pm0.020$ \\
\multicolumn{9}{c}{The first-harmonic quintuplet} \\
$2\nu-2\nur$            & $ 118\pm74$ & $0.986\pm0.100$ & $0.038\pm0.012$ & $0.030\pm0.051$ & $ 1.34\pm0.36$ & $0.272\pm0.043$ & $  3\pm17$ & $0.441\pm0.775$ \\
$2\nu- \nur$            & $  66\pm74$ & $0.808\pm0.178$ & $0.005\pm0.012$ & $0.920\pm0.386$ & $ 0.43\pm0.36$ & $0.552\pm0.131$ & $  2\pm17$ & $0.661\pm1.211$ \\
$2\nu      $            & $ 156\pm71$ & $0.950\pm0.072$ & $0.078\pm0.012$ & $0.996\pm0.026$ & $ 0.93\pm0.35$ & $0.391\pm0.060$ & $ 63\pm16$ & $0.956\pm0.042$ \\
$2\nu+ \nur$            & $ 108\pm74$ & $0.192\pm0.109$ & $0.006\pm0.012$ & $0.265\pm0.302$ & $ 0.52\pm0.36$ & $0.067\pm0.108$ & $  4\pm17$ & $0.235\pm0.598$ \\
$2\nu+2\nur$            & $ 193\pm74$ & $0.030\pm0.061$ & $0.044\pm0.012$ & $0.023\pm0.044$ & $ 1.50\pm0.36$ & $0.774\pm0.038$ & $ 35\pm17$ & $0.059\pm0.077$ \\
\hline
\end{tabular}
\end{table*}

\begin{figure*}[!t]
\figps{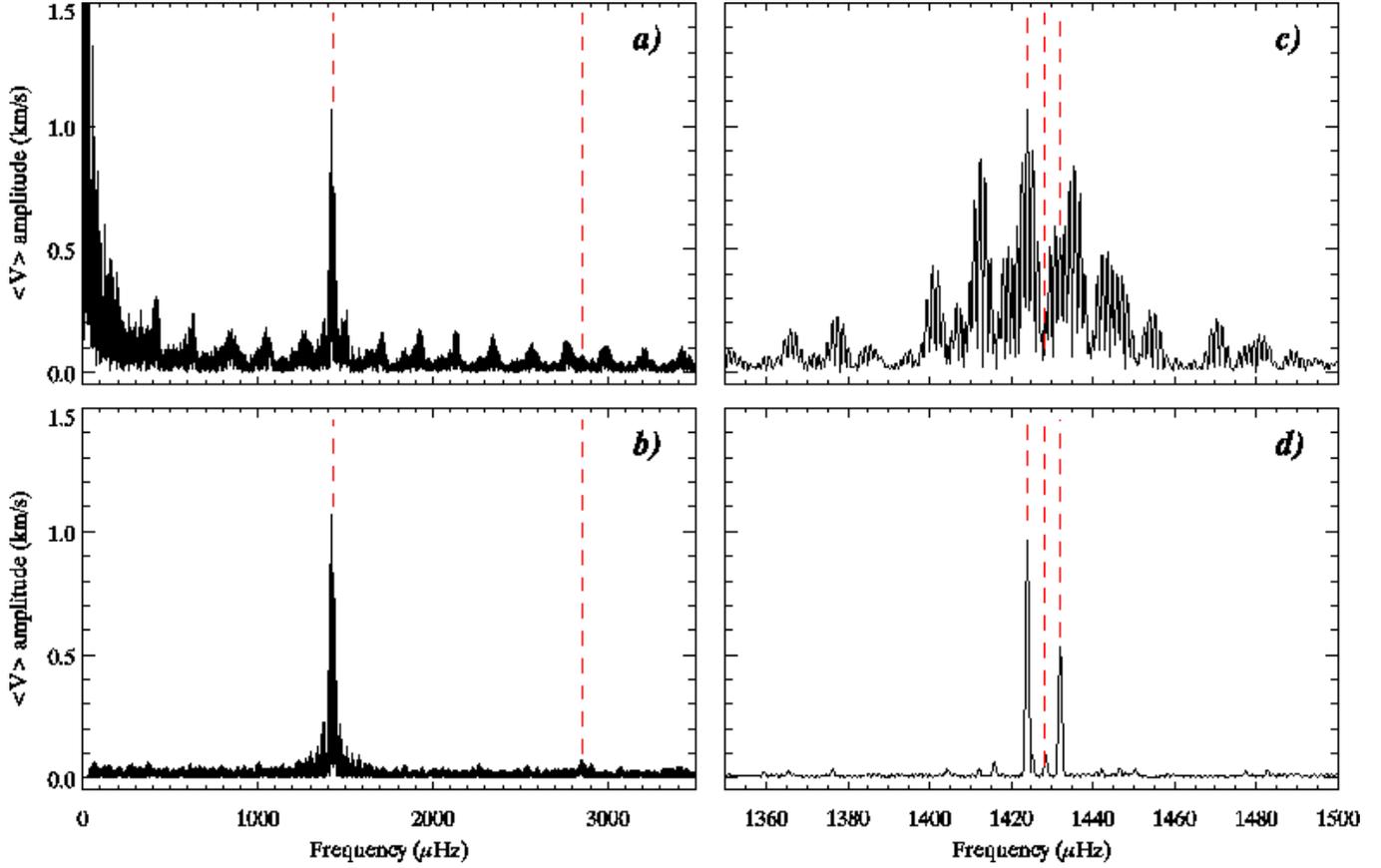}
\caption{Frequency analysis of the RV variation of the \ndlin\ line in the spectrum of \hr. 
The left panels show low-resolution amplitude spectra.
The vertical dashed lines mark the fundamental pulsation frequency ($\nu=1427.9958\,\mu$Hz) 
and its first harmonic ($2\nu=2855.916\,\mu$Hz).
{\bf a)} The amplitude spectrum computed from the raw radial velocities
shown in Fig.\,\ref{fig2}; {\bf b)} the amplitude spectrum after prewhitening
of the rotational modulation. The right panels present 
high-resolution amplitude spectra in the vicinity of the main pulsation frequency. 
The vertical dashed lines show the expected position of the triplet around the main frequency: 
$\nu-\nur, \nu, \nu+\nur$ (where $\nur=4058.265$\,nHz). {\bf c)} The amplitude spectrum after prewhitening 
of the rotational modulation; {\bf d)} the corresponding CLEANed amplitude spectrum.}
\label{fig3}
\end{figure*}

\begin{figure*}[!t]
\fifps{15cm}{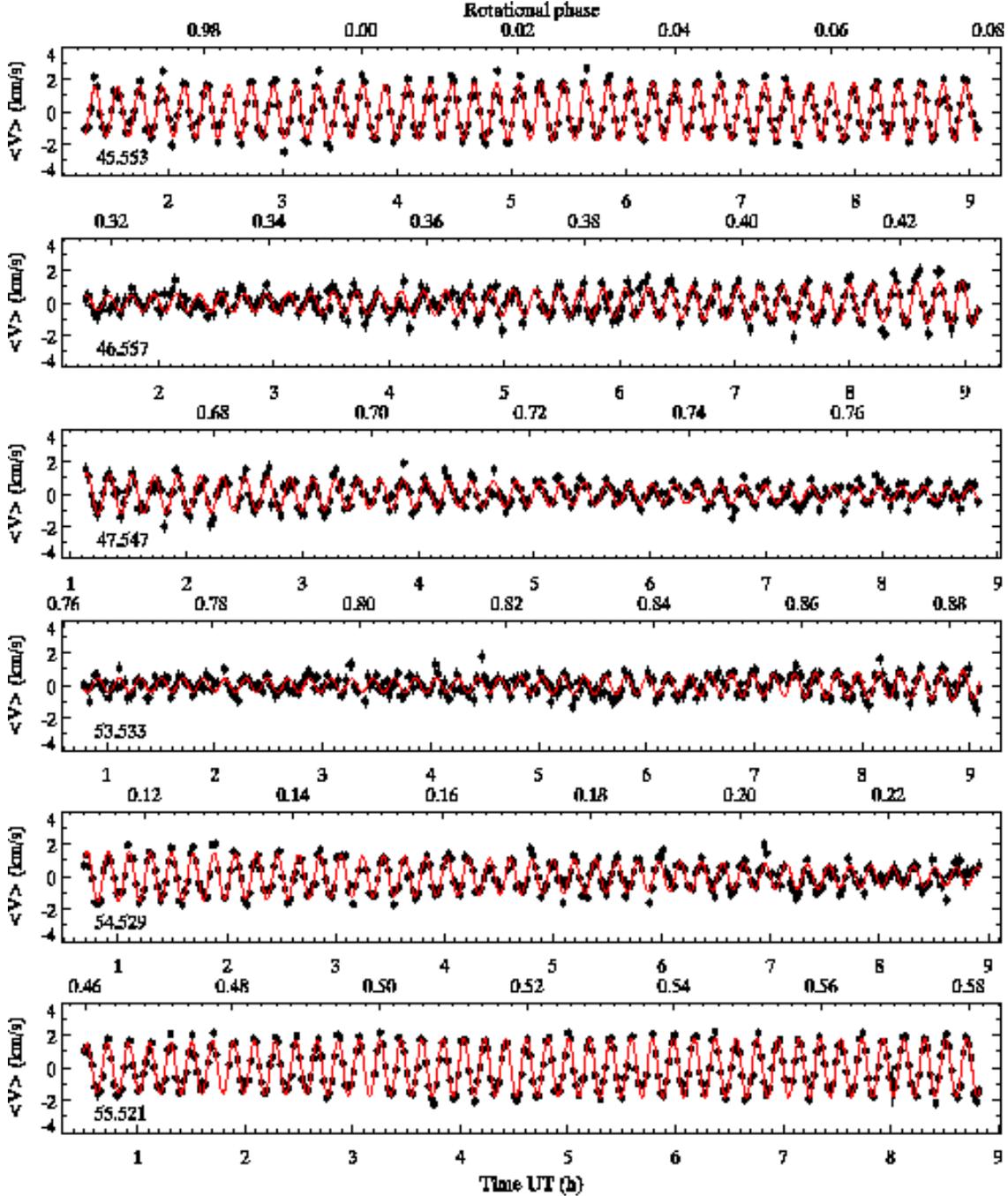}
\caption{Radial velocity curve for all observations of the \ndlin\ line in the spectrum of
\hr. Symbols show individual RV measurements, with the rotational modulation of RV removed 
using the Fourier fit illustrated in Fig.~\ref{fig2}. The solid curve shows RV
predicted by the frequency solution discussed in  the text.
Individual panels correspond to the data obtained during each of the six observing nights.
The number in the lower left of each panel gives HJD\,$-2451900$ of the start of
spectroscopic monitoring. The $x$-axis is labeled with UT (in hours) and rotation phases.}
\label{fig4}
\end{figure*}

\begin{figure*}[!t]
\fifps{\hsize}{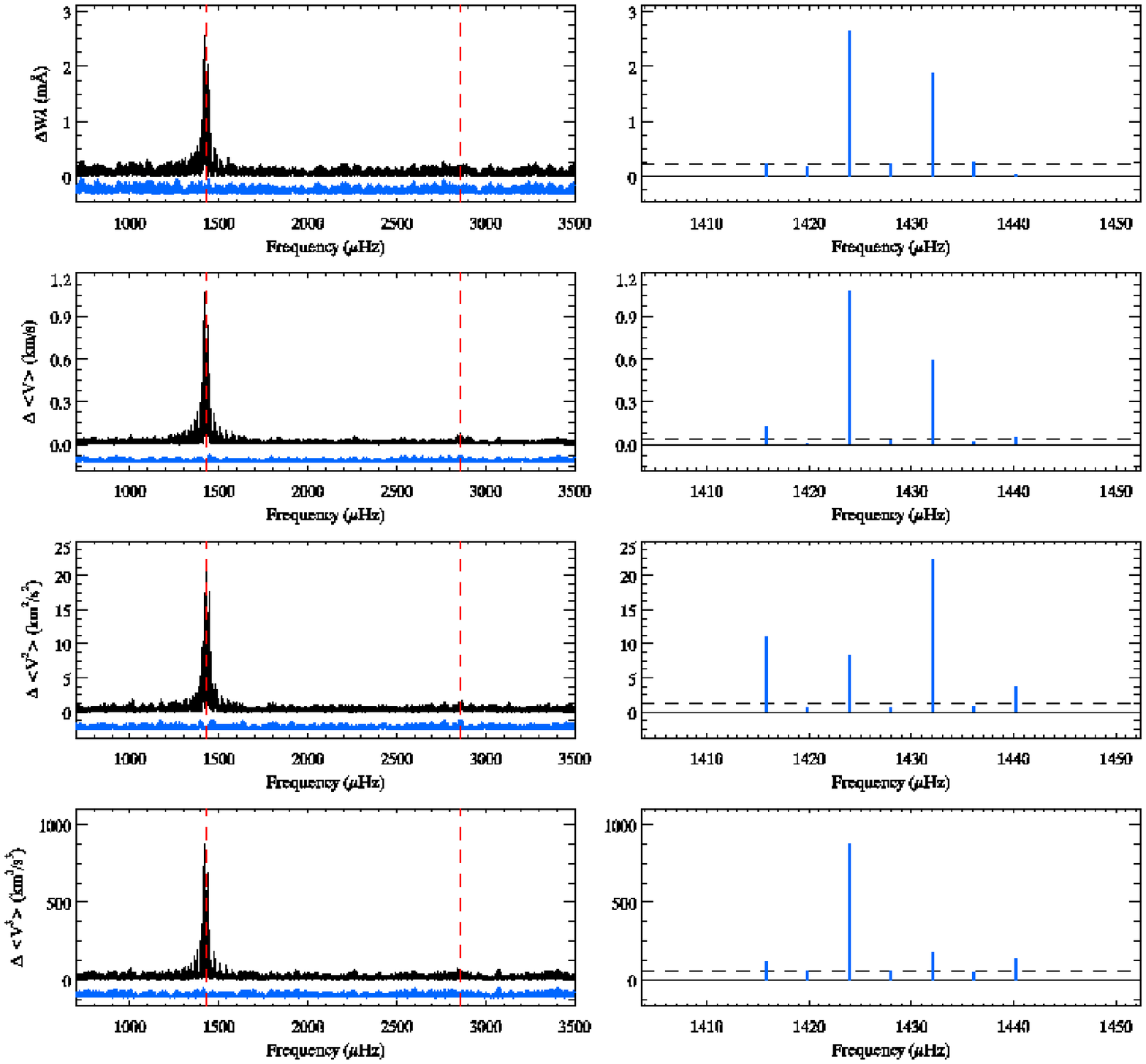}
\caption{Amplitude spectra of the pulsational variability of (in this order, from top to
bottom) the equivalent width \wl\ and the first three moments -- the radial velocity \va,
the second moment \vb, and the third moment \vc\ -- measured for the \ndlin\ line. The left panels
show low-resolution amplitude spectra. The vertical dashed lines marks the main
pulsation frequency and its first harmonic. The right sequence of panels illustrate 
amplitudes of the components of the fundamental septuplet ($\nu-3\nur, \ldots, \nu, \ldots,
\nu+3\nur$) derived with least-squares analysis. The
horizontal dashed line corresponds to the $3\sigma$ uncertainty of the fitted amplitudes. The
lower amplitude spectra in the left panels show the amplitude spectra after the frequency solutions
illustrated in the respective right panels (see also Table~\ref{tbl2}) have been removed from the data.
These residual amplitude spectra are shown at the same scale as the original spectra and
are shifted downward for display purpose.}
\label{fig5}
\end{figure*}

\begin{figure*}[!t]
\fifps{16cm}{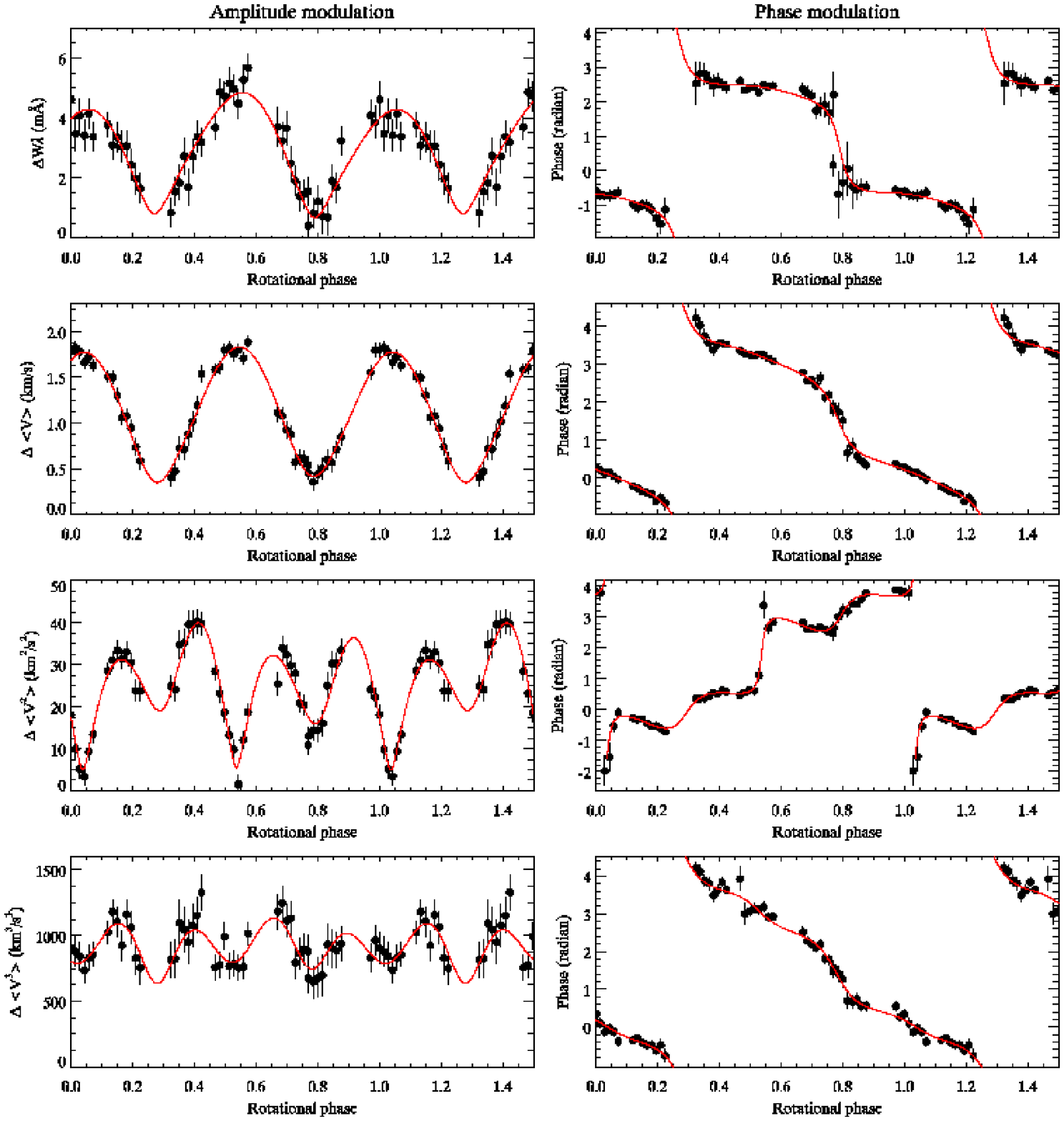}
\caption{Rotational modulation of the pulsation amplitude (left panels) and phase 
(right panels) of the variability of moments of the \ndlin\ line in the spectrum of
\hr. The panels show (in this order, from top to bottom) measurements of the equivalent width 
\wl, radial velocity \va, the second moment \vb, and the third moment \vc. Individual points in these
diagrams were calculated by fitting the main pulsation frequency 
($\nu=1427.9958\,\mu$Hz) to the subsets of 38 spectroscopic observations (about 5 pulsation
cycles or 56\,min). The smooth curves show rotational modulation expected from the frequency solution 
for the fundamental septuplet given in Table~\ref{tbl2} and illustrated in
Fig.~\ref{fig5}.}
\label{fig6}
\end{figure*}

A low angular degree of the non-radial mode excited in \hr\ is reflected in a fairly simple
pulsational line profile variation. Therefore, the integral observables, like the low-order line profile moments (Aerts et al.
\cite{APW92}), are sufficient for the purpose of frequency analysis and should 
yield essentially the same information as pixel-by-pixel line profile measurements (e.g.
Telting \& Schrijvers \cite{TS97}). We determined moments of the \ndlin\ line using
expressions given by Kochukhov \& Ryabchikova (\cite{KR01a}) whose formulas are equivalent to
the standard line profile moment definitions of Aerts et al. (\cite{APW92}). 

The equivalent width \wl, radial velocity \va, the second moment \vb\ (linked to the line
width), and the third moment \vc\ (a measure of the line asymmetry) were determined by
integrating normalized spectra in the fixed 2~\AA\ window centred at the \nd\ line. 
The relative precision of the RV measurements is considerably higher
compared to \wl\ and other moments. Consequently, we have chosen \va\ for detailed
frequency analysis. The 1860 RV determinations for the \nd\ line are illustrated in
Fig.\,\ref{fig2}. The data are phased with the rotational ephemeris 
\beq
HJD=2448312.23606+E\times2.851976
\eeq
(Kurtz et al. \cite{KWR97}) which are used throughout our
paper. An inhomogeneous Nd distribution causes a prominent double-wave rotational modulation
of RV. Pulsational variability is superimposed on these longer term changes and manifests
itself as a low amplitude high frequency scatter of the points in Fig.\,\ref{fig2}.  

In the frequency domain (Fig.\,\ref{fig3}a) a contribution of non-radial pulsation is clearly seen
as an excess of amplitude in the 1350--1500 $\mu$Hz frequency range. In this paper we are primarily
concerned with the pulsational changes of the spectrum of \hr\ and, hence, remove rotational
modulation of RV by subtracting a 5-order Fourier fit (the smooth curve in Fig.\,\ref{fig2}) from
the data in time domain. The amplitude spectrum after this prewhitening procedure is shown in
Fig.\,\ref{fig3}b. The high-amplitude peaks at low frequencies are gone and the pulsational
contribution to the RV variability stands out with an unprecedented clarity. Substantial pulsational
signature in the \va\ amplitude spectrum is observed only close to the main oscillation frequency of
\hr. This differs from the photometric pulsational behaviour  (Kurtz et al. \cite{KWR97}) where 
significant amplitude excess is also seen at the first and second harmonics.

The high-resolution amplitude spectrum (Fig.\,\ref{fig3}c) shows that aliasing remains a problem in
our single-site data despite a superb quality of the observational material. In a situation where
direct identification of the pulsation frequencies is hampered by the unfavourable window function, the
CLEAN technique (Roberts et al. \cite{RLD87}) often aids in distinguishing intrinsic
pulsation frequencies from their aliases. We applied the CLEAN algorithm to the \va\ measurements
using 200 iterations with gain=0.6. The resulting amplitude spectrum (Fig.\,\ref{fig3}d) shows two
high-amplitude peaks and a few weaker ones. Modification of the CLEANing parameters affects the
low-amplitude peaks but the two main frequencies remain unchanged and, thus, can be identified without
any ambiguity. Their position (respectively,  $1424.06$ and $1432.03$ $\mu$Hz) coincides with the
$\nu-\nur$ and $\nu+\nur$  variation seen in the high-speed photometry. This observation provides a
strong argument for the overall similarity of the frequency structure of the pulsational light and 
RV variation of \hr.

Given the satisfactory agreement between the main features of the spectroscopic and photometric
pulsational variability at the fundamental frequency of \hr, we proceed with the least-squares
analysis assuming that the oblique rotator model is valid for describing the \va\ variation and that
frequencies are always separated by multiples of $\nur$ from the fundamental frequency $\nu$ and its
harmonics. We use a non-linear least-squares algorithm to fit the whole fundamental septuplet and
the first-harmonic quintuplet. (It should be emphasized that the respective 12 frequencies 
do not belong to physically different non-radial modes but arise from the
rotational modulation of the same distorted oblique pulsation.) 
The radial velocity is approximated with the expression
\beq
\begin{array}{rl}
\langle V \rangle = & \sum\limits_{i=-3}^3 A_i \cos{\left[2 \pi \left((t-T_0) ( \nu+i\,\nur)+\varphi_i\right)\right]} + \\
                    & \sum\limits_{j=-2}^2 A_j \cos{\left[2 \pi \left((t-T_0) (2\nu+j\,\nur)+\varphi_j\right)\right]},
\end{array}
\label{eq1}
\eeq
where the epoch of phase zero is $T_0=2448312.23606$ and all amplitudes $A_{i,j}$, 
phases $\varphi_{i,j}$, and main pulsation frequency $\nu$ are
optimized simultaneously. The search for the best-fit rotation frequency can also be carried out.
The resulting value of the stellar rotation period,
$P_{\rm rot}=2.8624\pm0.0046$~d, agrees with $P_{\rm rot}=2.851976\pm0.00003$~d determined
by Kurtz et al. (\cite{KWR97}) but is certainly less precise due to a shorter time-span of our 
observations. Hence, we prefer to adopt the constant photometric $\nur=4058.265$~nHz in the fit with
Eq.\,(\ref{eq1}). The final parameters of the 12-frequency solution are reported in Table~\ref{tbl2}.
Some of the pulsational contributions have amplitudes below the formal detection threshold but are,
nevertheless, included here for completeness and to get an estimate of the upper limit of the RV
variability at the respective frequencies. On the other hand, the second-harmonic triplet was not
included in the fit because none of its components has amplitude above 16~\ms.

The fundamental pulsation frequency determined in the least-squares analysis is
$\nu=1427.9958\pm0.0072$~$\mu$Hz. This is the best determination of the main pulsation frequency of
\hr.  The comparison between observed RVs of the \nd\ line and prediction of the frequency solution
is illustrated in Fig.\,\ref{fig4}. Pulsational RV variation reaches 3.6~\kms\ peak-to-peak
amplitude which is the highest value ever observed for a roAp star. For all observing nights the stellar
pulsational behaviour is modelled with a good precision. The fundamental septuplet plus the
first-harmonic quintuplet explain 86\% of the data  variance. The residual amplitude spectrum of
\va\ is shown in the upper panel of Fig.\,\ref{fig5}. The average noise level is 14~\ms, whereas the
highest high-frequency residual peaks reach  $\approx$\,40~\ms. In the low-frequency domain we find
a 59~\ms\ peak at $\nu=10.239$~d$^{-1}$ and its alias at $\nu=11.238$~d$^{-1}$. Detection of this
variation is not statistically significant, but, if not caused by the instrumental effects, it can
be a candidate for a weak  $\delta$~Scuti-type pulsation mode.

Although we found a good overall concordance between the photometric and spectroscopic pulsation
behaviour, the amplitudes and phases of the individual frequency components turn out to be quite
dissimilar. The first striking discrepancy between pulsational light variation and RVs is a
significantly more sinusoidal behaviour of the latter observable. In fact, among the first-harmonic
quintuplet components only the $2\nu$ variation is detected reliably and its amplitude is 14 times
lower than the amplitude of the largest component of the fundamental septuplet. Compare this with
the time-resolved photometry in a Johnson $B$ filter (Kurtz et al. \cite{KWR97}) for which this
ratio is 5 and even the second harmonic can be detected. Another difference is a substantially weaker
main frequency component observed in the RV variation. A widely used diagnostic of the oblique
pulsator geometry $(A_{+1}+A_{-1})/A_0=\tan{i}\tan{\delta}$ (where $i$ is inclination angle the
rotation axis and  $\delta$ is the mode obliquity relative to this axis, see Kurtz \& Martinez
\cite{KM00}) is 9.3 according to Kurtz et al. (\cite{KWR97}) but this parameter is 66.9 for our
spectroscopic data, indicating that one or both characteristic angles are close to 90\degr.
The asymmetry between the amplitudes of the  $\nu\pm\nur$ pulsation components is also larger for
RV: $A_{-1}/A_{+1}=1.82$ versus 1.21 in photometry.

\begin{figure*}[!t]
\fifps{12cm}{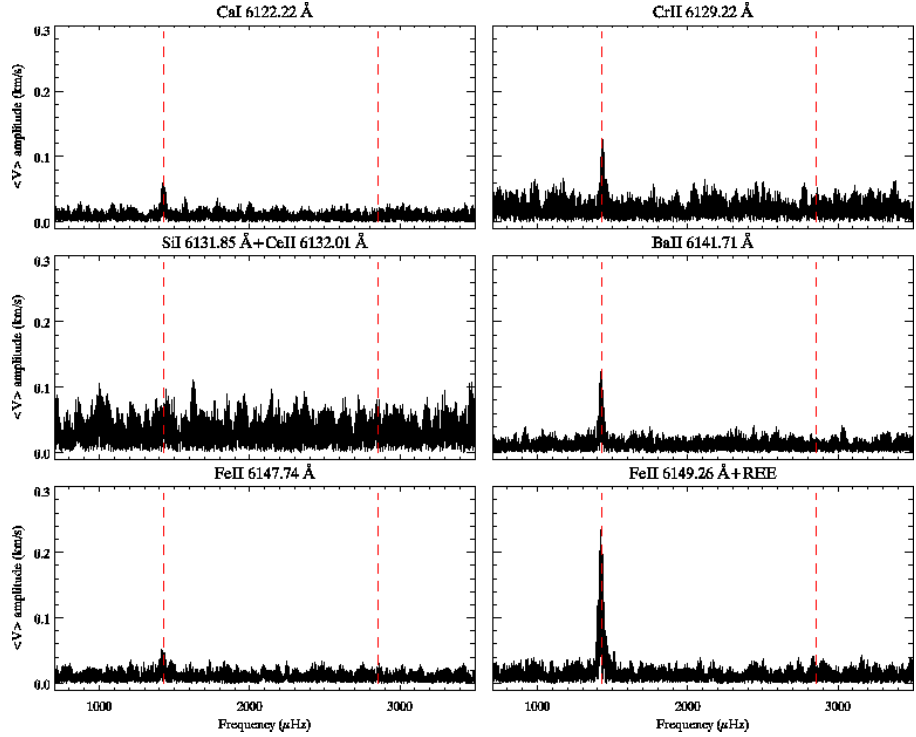}
\caption{Amplitude spectra of the RV variation observed for spectral features other than the \ndlin\
line. The vertical dashed lines mark the fundamental pulsation frequency ($\nu=1427.9958\,\mu$Hz) 
and its first harmonic ($2\nu=2855.916\,\mu$Hz).}
\label{fig7}
\end{figure*}

\subsection{Moment variation and rotational modulation}

We proceed with analysis of the pulsational variation of the equivalent width and higher-order
moments of the \ndlin\ line using approach similar to the one applied to RV oscillations. The
amplitudes and phases of the fundamental septuplet and first-harmonic quintuplet were optimized
simultaneously. The main pulsation frequency was kept constant at the value of
$\nu=1427.9958$~$\mu$Hz determined from the first moment. Results of the fit of \wl, \vb, and \vc\
are presented in Table~\ref{tbl2} and are illustrated in Fig.~\ref{fig5}.

It is clear that pulsational variation of all line profile observables considered here are detected
with extremely high $S/N$ at the main pulsation frequency. Weak harmonic variability is also found 
at $\approx$\,4$\sigma$ level for the \vb\ and \vc\ moments. Inspection of the residual amplitude
spectra of the line profile moments (Fig.~\ref{fig5}) does not reveal any additional statistically
significant amplitude peaks. Therefore, the frequency model defined by Eq.~(\ref{eq1}) is fully
adequate and explains all variability present in the time series of the line profile moments. 

An alternative presentation of the spectroscopic variability induced by the oblique non-radial
pulsation  in \hr\ can be obtained by considering pulsation with only the main frequency $\nu$ but
accounting for the rotational modulation by using a time-dependent amplitude and phase. This
formulation closely corresponds to the actual pulsation behaviour thought to be present in \hr\ and
is mathematically equivalent to fitting constant amplitudes and phases of the fundamental septuplet
components. Rotational modulation of the pulsations in \wl, \va, \vb, and \vc\ is illustrated in
Fig.~\ref{fig6}. The measurements shown in this figure were obtained by fitting cosine function to
47 subsets of our observations covering roughly 5 pulsation cycles. Rotational modulation of the
line profile characteristics is well defined and is in perfect agreement with the modulation
expected for the fundamental septuplet solution obtained above, confirming validity of both
approaches to time-series analysis of the spectroscopic observations of \hr.

The amplitude of \wl\ variation reaches $\approx$\,5~m\AA\ at the rotation phase of the magnetic
extrema. The phase modulation of the equivalent width changes is characterized by a sharper $\pi$
radian transition at the crossover phase compared to the \va\ phase modulation. Thus, \wl\ shows
more similarity to the photometric pulsational behaviour and, to some extent, can be used as a
proxy of the latter.

The radial velocity amplitude modulation is qualitatively similar to the \wl\ amplitude changes,
but does not show asymmetry between the two pulsation poles evident in the \wl\ data. We note that
amplitude changes of the RV variation found in our paper are consistent with the predictions of
the generic oblique pulsator model. On the other hand, we do not confirm a single-wave rotational
modulation of the REE RV amplitude claimed by Balona (\cite{B02}). Considering the large scatter of
the measurements and a poor rotation phase coverage evident in Fig.\,9 of Balona's paper, we
conclude that no rotational modulation of the RV oscillations was detected in his study and the
single-wave sinusoidal fit is not significant.

The second moment amplitude shows an unexpectedly complex rotational modulation with the two maxima and
two minima during each rotation cycle. The minima observed at the phases of maximum RV are deeper and
narrower compared to the decrease of \vb\ amplitude associated with the pulsational equator (rotation
phases $\varphi$\,$\approx$\,0.25 and 0.75). Below we will demonstrate that this interesting behaviour of
\vb\ is closely related to the distortion of oblique pulsation by the dipole component of the 
stellar magnetic field.

The amplitude maxima in the rotational modulation curves of \wl\ and \va\ presented in Fig.\,\ref{fig6}
occur slightly later than predicted by the ephemeris of Kurtz et al. (\cite{KWR97}). This
$\Delta\varphi=0.036\pm0.008$ discrepancy between the phases of the photometric and spectroscopic  pulsation
maxima may be intrinsic to the star or indicate an error in the adopted rotation period. If the latter is
true, $P_{\rm rot}=2.852055\pm0.000017$~d would bring in agreement the time of the photometric pulsation 
maximum (HJD=2448312.23606, Kurtz et al.) and our spectroscopic observations. The improved rotation period
deviates by less than 3$\sigma$ from the $P_{\rm rot}$ derived by Kurtz et al. All results presented in
our paper are based on the data covering just 3.6 stellar rotations and, therefore, are unaffected by
this small correction to the rotation period.  

\section{Pulsational changes in other lines}

\begin{table*}[!t]
\caption{Results of non-linear least-squares fit of the RV variation of spectral lines other than
\ndlin. \label{tbl3}}
\begin{tabular}{lrcrcrc}
\hline
\hline
Frequency  & \multicolumn{2}{c}{\ion{Ca}{i} 6122.22~\AA} & \multicolumn{2}{c}{\ion{Cr}{ii} 6129.22~\AA} & 
             \multicolumn{2}{c}{\ion{Si}{i} + \ion{Ce}{ii} 6131.9~\AA} \\
           & \multicolumn{1}{c}{$A$ (\ms)} & $\varphi$ & \multicolumn{1}{c}{$A$ (\ms)} & $\varphi$ & \multicolumn{1}{c}{$A$ (\ms)} & $\varphi$ \\
\hline
$\phantom{2}\nu-3\nur$  & $25\pm9$ & $0.511\pm0.055$ & $ 44\pm19$ & $0.509\pm0.068$ & $23\pm29$ & $0.484\pm0.206$ \\
$\phantom{2}\nu-2\nur$  & $12\pm8$ & $0.446\pm0.108$ & $ 24\pm18$ & $0.289\pm0.116$ & $45\pm27$ & $0.036\pm0.098$ \\
$\phantom{2}\nu- \nur$  & $62\pm8$ & $0.439\pm0.021$ & $ 67\pm17$ & $0.211\pm0.041$ & $49\pm27$ & $0.470\pm0.087$ \\
$\phantom{2}\nu      $  & $27\pm9$ & $0.095\pm0.051$ & $ 21\pm18$ & $0.807\pm0.143$ & $35\pm29$ & $0.189\pm0.131$ \\
$\phantom{2}\nu+ \nur$  & $53\pm8$ & $0.157\pm0.024$ & $115\pm17$ & $0.211\pm0.024$ & $81\pm27$ & $0.117\pm0.053$ \\
$\phantom{2}\nu+2\nur$  & $10\pm8$ & $0.407\pm0.133$ & $ 28\pm18$ & $0.674\pm0.100$ & $38\pm27$ & $0.288\pm0.113$ \\
$\phantom{2}\nu+3\nur$  & $13\pm9$ & $0.108\pm0.107$ & $ 43\pm19$ & $0.099\pm0.069$ & $51\pm29$ & $0.206\pm0.092$ \\
\hline
Frequency  & \multicolumn{2}{c}{\ion{Ba}{ii} 6141.71~\AA} & \multicolumn{2}{c}{\ion{Fe}{ii} 6147.74~\AA} & 
             \multicolumn{2}{c}{\ion{Fe}{ii} + REE 6149.26~\AA} \\
\hline
$\phantom{2}\nu-3\nur$  & $ 41\pm11$ & $0.326\pm0.042$ & $41\pm10$ & $0.536\pm0.040$ & $ 86\pm12$ & $0.392\pm0.021$\\
$\phantom{2}\nu-2\nur$  & $ 14\pm10$ & $0.239\pm0.122$ & $15\pm10$ & $0.591\pm0.105$ & $ 47\pm11$ & $0.144\pm0.037$\\
$\phantom{2}\nu- \nur$  & $132\pm10$ & $0.302\pm0.012$ & $45\pm~9$ & $0.573\pm0.033$ & $248\pm11$ & $0.287\pm0.007$\\
$\phantom{2}\nu      $  & $ 16\pm11$ & $0.018\pm0.109$ & $19\pm10$ & $0.895\pm0.085$ & $ 18\pm11$ & $0.779\pm0.103$\\
$\phantom{2}\nu+ \nur$  & $ 81\pm10$ & $0.173\pm0.020$ & $47\pm~9$ & $0.093\pm0.032$ & $146\pm11$ & $0.985\pm0.012$\\
$\phantom{2}\nu+2\nur$  & $ 26\pm10$ & $0.567\pm0.064$ & $ 4\pm10$ & $0.530\pm0.398$ & $ 60\pm11$ & $0.563\pm0.029$\\
$\phantom{2}\nu+3\nur$  & $  8\pm11$ & $0.269\pm0.210$ & $25\pm10$ & $0.205\pm0.064$ & $ 61\pm12$ & $0.007\pm0.030$\\
\hline
\end{tabular}
\end{table*}

Observations of \hr\ analysed in this paper for the first time permitted measurements of pulsation
amplitudes substantially below 50~\ms\ for individual spectral lines. Precision of our data enables
detection of pusational RV variation for essentally all strong and medium-strength lines.
Analyses of these absorption features was carried out using the same procedure as was outlined
above for the \ndlin\ line. Table~\ref{tbl3} reports amplitudes of the components of the
fundamental septuplet for 6 lines and blends in the 6120--6150~\AA\ spectral interval. Especially
important is the first definite detection and measurement of the pulsational RV variation for the
light and iron-peak elements, represented in the studied spectral region by the \ion{Ca}{i}
6122.22~\AA\ and  \ion{Fe}{ii} 6147.74~\AA\ lines. The substantially larger amplitude of pulsations
found for the \ion{Fe}{ii} 6149.26~\AA\ line reflects blending by an unidentified REE spectral line
(cf. Kochukhov \& Ryabchikova \cite{KR01a}). Oscillations are also readily detected for the
\ion{Ba}{ii} 6141.71~\AA\ line, whose pulsation amplitude in \hr\ is comparable to the pulsational
changes of this line in the roAp star \equ\ (Kochukhov \& Ryabchikova \cite{KR01a}).

Our measurements of the RV variation of metal lines (Table~\ref{tbl3}) and the \ndlin\ spectral
feature (Table~\ref{tbl2}) show a factor of 50 difference in amplitude and significant (sometimes
as large as  $\pi$ radian) discrepancy in phase. This picture is consistent with general
explanation of the pulsational spectroscopic behaviour of roAp stars proposed by Kochukhov \&
Ryabchikova (\cite{KR01a}) and  Ryabchikova et al. (\cite{RPK02}): the running pulsation wave
propagates outward through the chemically inhomogeneous stellar atmosphere with increasing
amplitude. Large difference in the pulsation characteristics of various lines arises from an
interplay between the depth-dependence of magnetoacoustic wave and chemical gradients formed by
radiative diffusion. In particular, the rare-earth elements are concentrated in high atmospheric
layers and probe regions with a large pulsation amplitude, whereas Ca and Fe lines originate from
the deeper layers where pulsations are very weak. The strong \ion{Ba}{ii} lines are probably
sensitive to the intermediate depths.

\section{Interpretation of moment variations}

\begin{figure*}[!t]
\fifps{15cm}{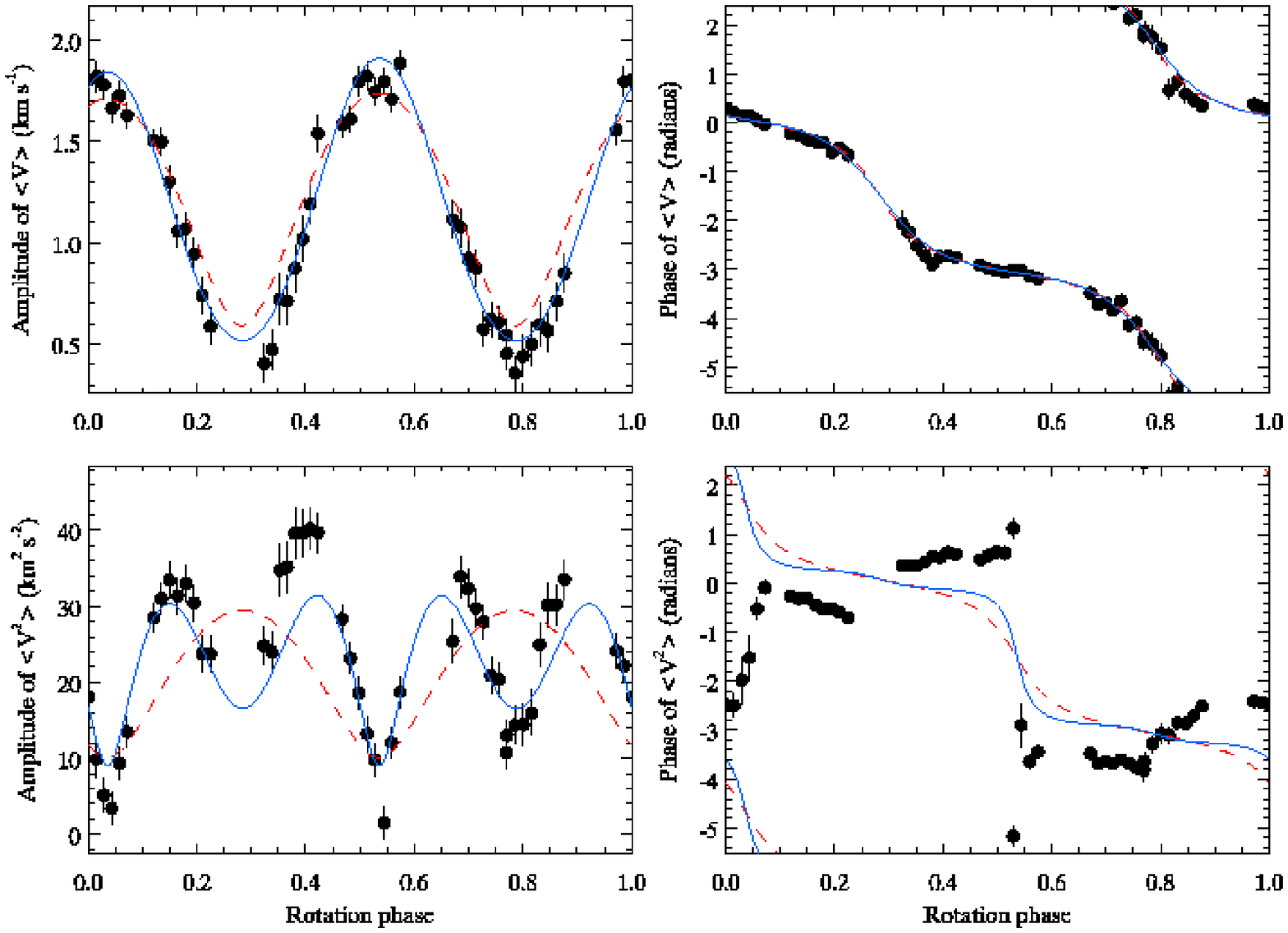}
\caption{Comparison between the observed rotational modulation of the pulsation amplitude
and phase of the \va\ and \vb\ moments of the \ndlin\ line (symbols) and 
predictions of the oblique 
pulsator model discussed in the text. The solid lines shows results obtained with the 
best-fit combination of the non-axisymmetric dipole ($\ell=1$, $m=-1,0,1$) and axisymmetric 
octupole ($\ell=3$, $m=0$) pulsation components, whereas the dashed line illustrates moment 
modulation expected from the dipole components alone.}
\label{fig8}
\end{figure*}

The problem of modelling variation of the line profile moments for stars pulsating in oblique
non-radial modes was addressed for the first time in a recent paper by Kochukhov (\cite{K05}). In
this study we derived relevant expressions and developed a numerical procedure suitable for 
analysis of the time-resolved spectroscopic observations of roAp stars. It was demonstrated that a
non-axisymmetric superposition of the pulsation and rotation velocity fields in oblique pulsators
leads to a qualitatively new behaviour of some of the line profile characteristics. Furthermore, a
distortion of the oblique dipole modes predicted by the recent theoretical studies of the  stellar
magnetoacoustic oscillations (Saio \& Gautschy \cite{SG04}; Saio \cite{S05}) can be readily
diagnosed through the moment analysis.  In particular, the shape of the pulsation phase  modulation
for the radial velocity and second moment is very sensitive to non-axisymmetric pulsation
components, whereas the rotational modulation of the second moment amplitude is best suited to
revealing axisymmetric magnetically  induced distortion of pulsations. The present paper deals with
the first application of the new moment technique to observations of a real roAp star and aims at
deriving a detailed model of the stellar  pulsation velocity field.

The moment code of Kochukhov (\cite{K05}) was employed to compute time-dependent amplitudes and
phases of \va\ and \vb\ for 100 equidistant rotation phases. The local line profiles were
approximated by a Gaussian with $\sigma=5$~\kms. Inclination angle $i=68.2\degr$ and
projected rotation velocity $v_{\rm e}\sin i=33.0$~\kms\ were adopted following results of the model
atmosphere and surface structure analysis of \hr\ presented by Kochukhov et al. (\cite{KDP04}).
The free parameters of our spectroscopic oblique pulsator model included the obliquity $\delta$ and
phase angle $\chi$ giving orientation of the pulsation axis and amplitudes of the $\ell=1$,
$m=0, \pm1$ and $\ell=3$, $m=0$  pulsation components. These parameters were optimized through the
comparison of the model predictions with the observed rotational modulation of the \va\ and \vb\
amplitudes and the first moment phase. An equal weight was given to all three observables and
horizontal pulsation motions were neglected. 

The final set of parameters of the oblique pulsator model required to fit variation of the \ndlin\
line profile moments is summarized in Table~\ref{tbl4}. Comparison between observations and model
predictions is illustrated in Fig.\,\ref{fig8}. All considered pulsation observables except the
phase of \vb\ are either reproduced successfully (\va\ amplitude and phase) or fairly well (\vb\
amplitude) by our model. It appears that complex  modulation of the second moment amplitude is
explained by the presence of $\ell=3$, $m=0$ pulsation. It definitely cannot be reproduced by any
combination of the dipole pulsation components (see Fig.\,\ref{fig8}) alone. We find that at the
formation depth of the \ndlin\ line pulsation geometry is slightly non-axisymmetric, with the
amplitude of the $\ell=1$, $m=\pm1$ components not exceeding $\sim$\,15\% of the contribution due
to the $\ell=1$, $m=0$ spherical harmonic. On the other hand, we uncover a prominent distortion of
the dipole pulsation in \hr\ by the global magnetic field. It manifests itself through the presence
of a substantial axisymmetric octupole ($\ell=3$, $m=0$) pulsation component and leads to
enhancement of velocity amplitude close to the  pulsation poles. These findings are in good
agreement with the outcome of pulsation Doppler imaging of \hr\ (Kochukhov \cite{K04b}) and are
in line with theoretical predictions of Saio \& Gautschy (\cite{SG04}) and Saio (\cite{S05}).

Precise information about orientation of the pulsation axis obtained in our study demonstrates
that oblique oscillations in \hr\ are aligned with the dipolar magnetic field component. The latest
and most detailed modelling of the magnetic topology of \hr\ (Kochukhov et al. \cite{KDP04}) has
yielded magnetic obliquity $\beta=86.8\pm6.2$\degr\ which coincides with the pulsation mode obliquity
$\delta=88.5\pm0.6$\degr\ determined through the moment analysis.

The present model describing variation of the \ndlin\ line cannot be improved by introducing
high-order non-axisymmetric pulsation contributions. A formal least-squares analysis also yields the 
ratio of the horizontal to vertical pulsation amplitude, $K$, indistinguishable from zero. 
However, the latter result may reflect a failure
of the standard constant-$K$ parameterization of the horizontal pulsation motions adopted in the moment
technique. In fact, this approximation is poorly suited to characterize latitude-dependent horizontal 
pulsation amplitude expected for magnetoacoustic pulsators (Saio \& Gautschy \cite{SG04}).

\begin{table}[!t]
\caption{Parameters of the oblique pulsator model of \hr\ determined from
the variation of the \ndlin\ line profile moments. This table reports stellar inclination
$i$, angles $\delta$ and $\chi$ specifying orientation of the pulsation axis, and
amplitudes (renormalized to yield the maximum surface velocity) of different spherical harmonic 
contributions to the pulsation velocity field. \label{tbl4}}
\begin{tabular}{cc}
\hline
\hline
Parameter & Value \\
\hline 
$i$ (\degr) & 68.2 \\
$\delta$ (\degr) & $88.5\pm0.6$\\
$\chi$ (\degr) & $167.1\pm0.5$ \\
$A(\ell=1,m=0)$ (\kms) & $3.33\pm0.03$ \\
$A(\ell=1,m=\pm1)$ (\kms) & $0.51\pm0.02$ \\
$A(\ell=3,m=0)$ (\kms) & $2.90\pm0.11$ \\
\hline
\end{tabular}
\end{table}

The most significant limitation of the modelling described in this Section is
related to the fact that we disregard inhomogeneous distribution of Nd at the surface of \hr.
Neodymium surface map reconstructed by Kochukhov et al. (\cite{KDP04}) shows relative 
underabundance of this element along the magnetic equator. Consequently, the moment diagnostic 
is somewhat less sensitive to the pulsation wave properties in these regions of the
stellar surface. However, we do not expect results of the overall moment analysis to be
significantly distorted by the presence of Nd spots because in our modelling information is mostly extracted
from the high-amplitude LPV observed at the rotation phases when no large Nd abundance gradients are present
at the visible stellar disk.

\begin{figure*}[!t]
\fifps{15cm}{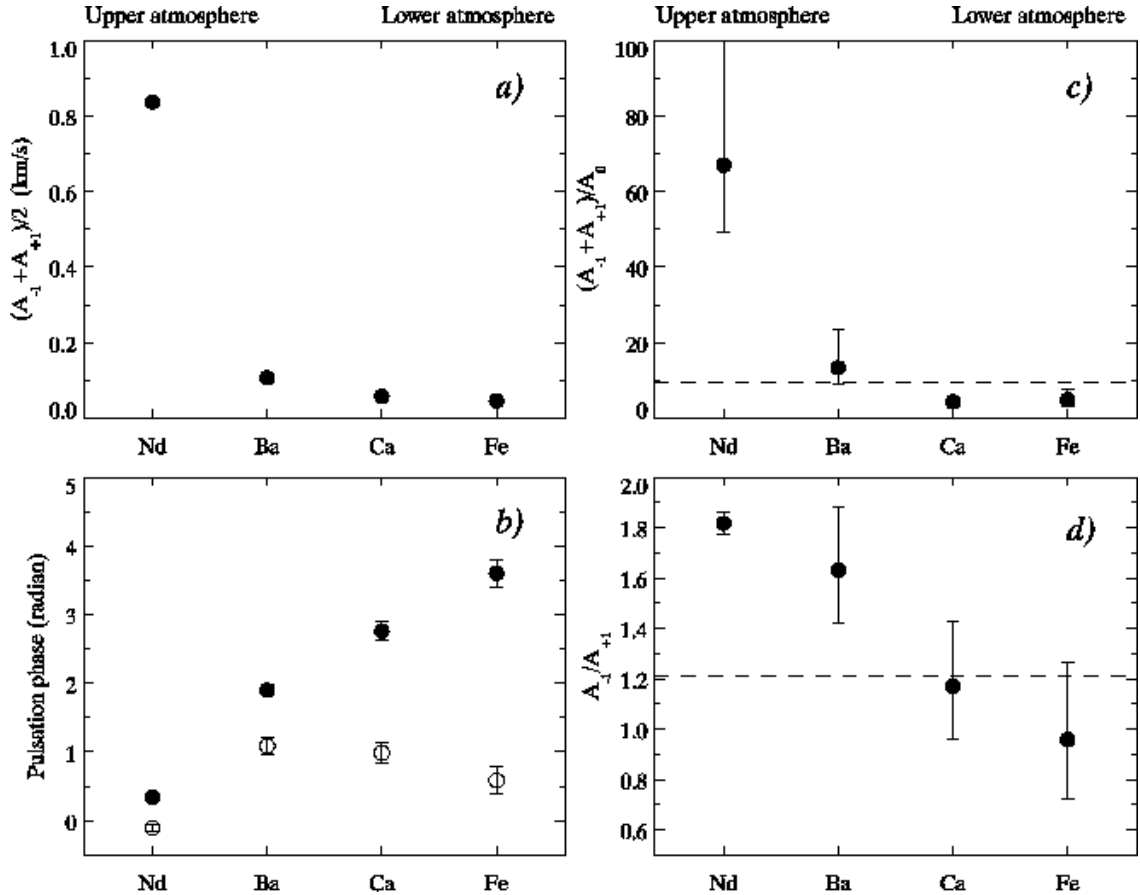}
\caption{Depth dependence of the magnetoacoustic wave properties in the atmosphere
of \hr. {\bf a)} $(A_{-1}+A_{+1})/2$ characterizes variation with depth of the average amplitude of the $\nu\pm\nur$ 
components, {\bf d)} depth dependence of the respective phases of the $\nu+\nur$ (open circules) 
and $\nu-\nur$ (filled circles) pulsation,
{\bf c)} the $(A_{-1}+A_{+1})/A_0$ parameter characterizes strength of the main 
frequency component relative to the $\nu\pm\nur$ components, {\bf d)} $A_{-1}/A_{+1}$ measures
asymmetry of the $\nu\pm\nur$ components. The symbols show results obtained for individual spectral 
lines, whereas the dashed line corresponds to the photometric pulsation behaviour in a Johnson
$B$ filter.}
\label{fig9}
\end{figure*}

\section{Discussion and conclusions}

Despite a long history of the high-speed photometric observations of pulsations in roAp stars, mechanisms
responsible for the pulsational luminosity variation are not understood. In particular, no 
theoretical model has been able to explain the observed complexity  (non-linearity, peculiar
wavelength dependence, etc.) of the photometric time series. Thus, no simple relation can be
established between changes of luminosity and fundamental pulsation characteristics,
such as the pulsation displacement vector. This is why all interpretations of the photometric pulsational
variability of roAp stars start with the \textit{assumption} that geometrical picture of the relative
luminosity changes, $\delta I/I$, is the same as for the vertical component of the displacement vector and
is given by a sum over spherical harmonic functions. Thus, even the latest state-of-the-art
theoretical investigations into pulsational light variability (e.g. Bigot \& Dziembowski \cite{BD02}) have
to abandon hopes to build a physically consistent picture and present essentially a \textit{geometrical}
model of the continuum intensity variation associated with {\it p-}mode oscillations. A progress in this
field requires detailed non-adiabatic MHD modelling (first models of this kind are being developed by 
Saio (\cite{S05})) as a basis for estimating contributions of different terms to the observed luminosity
variation of roAp stars. Nevertheless, the problem of a low intrinsic diagnostic content of a disk-integrated
photometric observable (cf. Saio \& Gautschy \cite{SG04}) could never be overcome.

In contrast to unsatisfactory understanding and limited information content of the pulsational light
curves, spectroscopic observations of rapid oscillations permit a straightforward astrophysical interpretation.
The observed pulsation velocity fluctuation is directly related to the time depedence of the pulsation
displacement vector. Given the absence of significant pulsational changes of the strength or equivalent
width of diagnostic lines, we can also assert that the observed RV variation is not visibly influenced by
secondary effects, such as rapid changes of temperature and density of atmospheric plasma. Consequently,
interpretation of the spectroscopic observations of roAp pulsations offers us a real opportunity to develop 
\textit{physical} models of magnetoacoustic pulsation for the first time since the discovery of roAp stars.
Furthermore, the presence of strong vertical gradients of chemical composition in the atmospheres of roAp
stars enables an unprecedented vertical resolution of pulsation waves. This makes spectroscopic studies of roAp
pulsations a very promising research direction, both in terms of its enourmous diagnostic potential and 
possibility of establishing a physically realistic model of the pulsational phenomena from time-resolved
observations. 

The line profile analysis has become a standard tool for the investigation into pulsational variability of
many types of oscillating stars. Yet, due to their short pulsation periods, roAp stars were not studied
before with this technique. The present report on LPV in the prototype roAp star
\hr\ fills this gap and represents a significant step towards developing observational methods and
analytical tools able to bring us closer to understanding intricate nature of the magnetoacoustic pulsations
in magnetic A stars.

Our line profile study of \hr\ focuses on the variability of the \ndlin\ line, which shows maximum amplitude
of the pulsational variability among spectral features in the studied wavelength region. It also exhibits one
of the largest pulsation amplitudes ever observed in a roAp star. Spectroscopic material described in our
paper shows unambigous evidence of the pulsational line profile variability and presents the very first
comprehensive study of this phenomenon in a star pulsating in oblique non-radial mode. In particular, it
enabled us to detect and characterize rotational modulation of the LPV pattern -- an effect 
expected for roAp stars long ago (e.g. Baade \& Weiss \cite{BW87}) but never observed until now.

Pulsational changes of the equivalent width and the first three moments of the \nd\ line are detected with a
very high $S/N$ ratio and studied over the whole stellar rotation cycle. Unlike strongly non-sinusoidal
pulsation light curves (Kurtz et al. \cite{KWR97}), all spectroscopic observables show only a marginal
presence of the first and no signature of the second harmonic in their amplitude spectra.
Following Balona (\cite{B02}), we interpret this observation to be an evidence that spectroscopy probes
sinusoidal pulsation of the star, whereas variation of temperature and other parameters contributing to
luminosity fluctuation is related to the pulsation displacement in a complex non-linear way.

Time-series analysis of the RV variation resulted in a remarkably accurate determination of the main
pulsation frequency of \hr\ and possible improvement of the rotation period of the star. 

Taking advantage of the comprehensive observation of the LPV in the \ndlin\ line, we develop a physically
consistent oblique pulsator model of \hr. This model directly characterizes horizontal geometry and
amplitude of the pulsation displacement in the upper atmospheric layers where strong REE lines are formed.
Rapid variation of the first two line profile moments is interpreted using a new version of the moment
mode identification method (Kochukhov \cite{K05}). We derive orientation of the pulsation axis and
amplitudes of various spherical harmonic contributions to the pulsation geometry. These results generally
agree with the pulsation Doppler imaging map of velocity field obtained for \hr\ in our earlier study 
(Kochukhov \cite{K04b}). 

Using the first detailed map of pulsations in a roAp star we are able to test predictions of recent
theoretical studies of magnetoacoustic pulsations. Two main types of the distortion of the basic oblique
dipole pulsator geometry were suggested. On one hand, Bigot \& Dziembowski (\cite{BD02}) take into account the
interaction between weak magnetic field, pulsation, and rotation and predict a non-axisymmetric dipole
pulsation geometry, which is not aligned with the magnetic field axis. On the other hand, Saio \& Gautschy
(\cite{SG04}) and Saio (\cite{S05}) neglect the stellar rotation and instead suggest that the dominant
distortion of pulsation appears due to the strong magnetic field influence. Such a distortion manifests itself
in the presence of substantial odd-$\ell$ harmonic components in the pulsation geometry. Our moment analysis
results show a small (but significant) non-axisymmetric contribution, which could be ascribed to the $\ell=1$,
$m=\pm1$ components, but we find no misalignment of the pulsation and magnetic field axes suggested by Bigot
\& Dziembowski (\cite{BD02}) for \hr. At the same time, we reveal a strong $\ell=3$, $m=0$ pulsation
component, in a good agreement with the model developed by Saio \& Gautschy (\cite{SG04}). This demonstrates
that magnetic distortion of pulsation modes is by far more important effect compared to the rotation and
magnetic field coupling investigated by Bigot \& Dziembowski (\cite{BD02}). Although a future ultimate model
of roAp pulsations should certainly include effects scrutinized by Bigot \& Dziembowski, neglect of the most
important magnetic perturbation of {\it p-}modes in their modified oblique pulsator theory makes it
inapplicable to roAp stars with the average magnetic field strength above $\approx$\,1~kG, which includes \hr.
On the other hand, success of the theoretical framework proposed by Saio \& Gautschy (\cite{SG04})
demonstrates that their model is based on a realistic description of the physics of magnetoacoustic pulsation
and, hence, represents a major progress in understanding excitation and propagation of pulsation waves in roAp
stars.

Unprecedented precision of the time-series spectroscopic observations of \hr\ analysed here permitted the
first definite detection of weak pulsational RV changes in the lines of light and iron-peak elements. This makes
it possible to attempt a tomographic mapping (Kochukhov \cite{K03}) of the vertical structure of the pulsation mode
in \hr. Such a procedure requires accurate information about formation depths of the spectral lines involved
in the analysis. Chemical stratification significantly affects the depths from which lines of individual ions
originate and this is why a model atmosphere analysis and stratification diagnostic are necessary prerequisites
to reconstructing a realistic picture of the depth dependence of pulsation characteristics. Such a study lies outside
the scope of this paper and cannot be performed using spectra with a limited wavelength coverage available to us.
Nevertheless, it is possible to probe certain general properties of the pulsation wave structure by using results of
the tomographic mapping of pulsations in \equ\ (Ryabchikova et al. \cite{RPK02}), which was largely based on the same
set of spectral lines as analysed here. Similarity of the atmospheric parameters and chemical composition of \hr\
and \equ\ suggests that the vertical abundance profiles and line formation depths are comparable in the two stars.
Thus, we can assert that the \ndlin\ line is formed in a high atmospheric layer, the \ion{Ba}{ii} 6141.71~\AA\
line is sensitive to intermediate depths, whereas lines of Ca and Fe should originate from deeper layers. At
the same time, the line of \ion{Ca}{i} 6122.22~\AA\ investigated here is considerably stronger that \ion{Fe}{ii}
6147.71~\AA\ and, hence, it should form somewhat higher compared to the iron line. Using this tentative
assignment of the relative formation depths of the four lines we are able to obtain the first picture of the
vertical cross-section of the pulsation mode in \hr. Information about depth-dependent behaviour of the
amplitudes and phases of the fundamental triplet is summarized in Fig.\,\ref{fig9}.

Our observations establish that the amplitude of pulsations increases by a factor of 17 from the lower to upper
layers in the atmosphere of \hr\ (Fig\,\ref{fig9}a). Such a behaviour is an expected consequence of the constancy
(or only a small change) of the pulsation energy density $\rho v^2/2$ in the presence of a rapid decrease
of the gas density in stellar astmosphere. Large pulsation phase difference between the RV variations of
spectral lines sensitive to upper and lower atmospheric layers clearly points to a short vertical length of
pulsation waves. Comparison of the pulsation phases derived for different lines in the studied spectral region of
\hr\ does not reveal a bimodal phase distribution but shows a quasi-continuous phase change with depth
(Fig\,\ref{fig9}b). Therefore, we conclude that the vertical structure of pulsation in the atmosphere of \hr\
resembles a running wave rather than a standing wave. With the data available here, we find no evidence for the 
existence of pulsation node in the atmosphere of \hr. 

Comparison of our results with the theoretical non-adiabatic calculation (Saio \cite{S05}) shows a reasonable
agreement with respect to the magnitude of the pulsation amplitude increase towards the upper atmospheric layers
and the absence of pulsation node in the line-forming region. However, a significant depth dependence of pulsation
phase is not expected for this theoretical model.

In addition to probing depth dependence of the average pulsation amplitude and phase, our measurements of the RV
variation of the lines of different ions enable extraction of the information about changes of the pulsation wave
geometry with height. A common procedure to characterize oblique pulsator geometry and its distortion by the effects
of stellar rotation is to study asymmetry of the $\nu\pm\nur$ peaks and estimate relative amplitude of the variation
at the unshifted frequency $\nu$. The respective observables, $A_{-1}/A_{+1}$ and $(A_{-1}+A_{+1})/A_0$, derived from
the light curves of roAp stars have been interpreted in a number of theoretical investigations (see Kurtz \& Martinez
\cite{KM00} for recent overview) under the assumption that these parameters permit a simple geometrical
interpretation within the oblique pulsator framework. Our study of \hr\ reveals that, similar to a dramatic depth
dependence of the pulsation amplitude and phase, relative amplitudes of the triplet components do not have 
well-defined values consistent with those inferred from the photometric observations of \hr\ but change significantly
with height in the stellar atmosphere. In particular, analysis of the \ion{Fe}{ii} and \ion{Ca}{i} lines shows that
pulsation in the lower part of the photosphere has a moderate asymmetry of the $\nu\pm\nur$ components and noticeable
contribution of the pulsation at the main frequency $\nu$. Both characteristics agree with photometric results (Kurtz
et al. \cite{KWR97}). However, asymmetry of the rotationally separated peaks increases and pulsation at the unshifted
frequency decreases below the formal detection threshold when the diagnostic \nd\ line sensitive to the upper
atmospheric layers is considered (see Fig.\,\ref{fig9}c, d). 

Observed pulsation behaviour of different lines in the \hr\ spectrum can be attributed to a signature of  the depth dependent
superposition of the acoustic and magnetic pulsation wave components. Importance of the latter contribution
presumably increases with height following the corresponding increase of the ratio of magnetic to gas energy density.
Furthermore, in the lower part of atmosphere dissipation of the magnetic slow waves and related horizontal pulsation
motions (Saio \& Gautschy \cite{SG04}) could also contribute to the observed spectroscopic and photometric pulsation
variability, introducing further difference with oscillation observed in the upper atmosphere using REE lines. Thus,
a common notion of $A_{-1}/A_{+1}$ and $(A_{-1}+A_{+1})/A_0$ as ``geometrical'' parameters typical of the whole
pulsation structure is meaningless. Both quantities show significant variation with depth and should always be
associated with specific layers of the atmosphere of a pulsating magnetic star. Thus, extreme caution is called for a purely
geometrical interpretation of the relative amplitudes of the $\nu, \nu\pm\nur$ components. 

Interpretation of the unique high time resolution spectroscopic data presented here inevitably calls for a significantly
more complex three-dimensional picture of the structure of pulsating cavity in roAp stars compared to the plane-parallel
view provided by the traditional oblique pulsator paradigm. There is clearly a need to advance from a rudimentary
phenomenological model to a self-consistent physical description of the propagation of magnetoacoustic waves in stellar
atmospheres. The first important step in this direction is self-consistent modelling of the vertical and horizontal 
chemical inhomogeneities and pulsational variability using pulsation Doppler Imaging method (Kochukhov \cite{K04a,K04b}).
This modelling can be applied to spectral lines formed at different optical depths in order to deduce a 3-D structure
of pulsation velocity field. This modelling is currently underway for \hr\ and will be presented in a forthcoming publication.

\begin{acknowledgements}
This study was partially supported by the post-doctoral stipend from the Swedish 
Research Fund and by the Lise Meitner fellowship from the Austrian Science Fund 
(FWF, project No. M757-N02).
\end{acknowledgements}

\end{document}